\documentclass[a4paper,12pt,oneside]{article}

\usepackage{etex}
\reserveinserts{28}
\usepackage[ngerman, english]{babel}		
\usepackage[affil-it]{authblk}
\usepackage{amsmath}
\usepackage{amssymb}
\usepackage{amsfonts}
\usepackage{arrayjobx}
\usepackage{arydshln}
\usepackage{bbding}
\usepackage{calc}
\usepackage[format=default,font=footnotesize,singlelinecheck=false,labelfont=bf]{caption}
\usepackage{cancel}
\usepackage{chngcntr}					
\usepackage[bulletsep]{collref}
\usepackage[usenames,dvipsnames]{color}	
\usepackage{colortbl}
\usepackage{courier}
\usepackage{dsfont}
\usepackage[ntheorem]{empheq} 			
\usepackage{exscale}             			
\usepackage{feynmp}
\usepackage{flafter}
\usepackage[nomessages]{fp}
\usepackage[latin1,utf8x]{inputenc}		
\usepackage{lipsum}
\usepackage{lmodern}						
\usepackage{mathdots}					
\usepackage{mathrsfs}
\usepackage{multicol}
\usepackage{multirow}
\usepackage{nicefrac}
\usepackage[amsmath,amsthm,framed,hyperref,thref]{ntheorem}
\usepackage{pdfpages}					
\usepackage{pict2e}
\usepackage{pifont}
\usepackage{pstricks}					
\usepackage{pst-node}					
\usepackage{oldgerm}
\usepackage{rotating}
\usepackage{scalerel}
\usepackage{scrpage2}
\usepackage{soul}
\usepackage{tabularx}
\usepackage{tensor}
\usepackage{todonotes}					
\usepackage{tocloft}
\usepackage{url}
\usepackage[]{units}
\usepackage{verbatim} 					
\usepackage{wrapfig}		 
\usepackage{xcolor}						
\usepackage{xspace} 						

\usepackage{microtype}					

\usepackage{tikz}
\usetikzlibrary{arrows, calc, matrix, positioning, shapes}
\tikzset{
	>=stealth',
	punkt/.style={
		minimum height=2em,
		text centered},
	pil/.style={
		->,
		thick,
		shorten <=2pt,
		shorten >=2pt,}
}

\newlength{\armlen}

\allowdisplaybreaks[1]

\def\ps@plain{\let\@mkboth\@gobbletwo%
     \let\@oddfoot\@empty\def\@oddhead{\reset@font\hfil\thepage}%
     \let\@evenfoot\@empty\def\@evenhead{\reset@font\thepage\hfil}}

\usepackage[plainpages=false]{hyperref}	
\hypersetup{pdfborder = 0 0 0,			
			bookmarksopen = true,		
			bookmarksnumbered = true,	
			}
\allowdisplaybreaks              

\theoremstyle{marginbreak} 
\theoremheaderfont{\normalfont\bfseries}
\theorembodyfont{\itshape}
\theoremseparator{}
\theorempreskipamount\topsep
\theorempostskipamount\topsep
\theoremnumbering{arabic}
\theoremsymbol{}

\newcommand{\say}{\raise0.45mm\hbox{\tiny$\Box$\hspace{0.5cm}}}

\newcommand{\nn}{\nonumber}

\makeatletter
\long\def\@makefntext#1{\parindent 1em\noindent\hbox to 2em{}\llap{\@thefnmark\phantom{.}\;}#1}
\makeatother
\makeatletter
\newcommand{\subalign}[1]{%
  \vcenter{%
    \Let@ \restore@math@cr \default@tag
    \baselineskip\fontdimen10 \scriptfont\tw@
    \advance\baselineskip\fontdimen12 \scriptfont\tw@
    \lineskip\thr@@\fontdimen8 \scriptfont\thr@@
    \lineskiplimit\lineskip
    \ialign{\hfil$\m@th\scriptstyle##$&$\m@th\scriptstyle{}##$\crcr
      #1\crcr
    }%
  }
}
\makeatother

\newcommand{\dAl}{\ensuremath{\Box}}

\newcommand{\G}{\mathfrak{G}}

\DeclareMathOperator{\dd}{d\!}
\newcommand{\DD}[1][]{D^{#1}\!}

\DeclareMathOperator{\hodge}{\raise-1mm\hbox{\ensuremath{*}}}

\DeclareMathOperator{\T}{T}

\newcommand{\fund}{\raise-0.35mm\hbox{$\Box$}}

\newcommand{\Complex}{\ensuremath{\mathbb{C}}}

\newcommand{\Spin}{\ensuremath{\mathbb{S}}}
\newcommand{\Proj}{\ensuremath{\mathbb{P}}}
\newcommand{\CP}{\ensuremath{\mathbb{CP}}}
\newcommand{\Tw}{\ensuremath{\mathbb{T}}}
\newcommand{\PT}{\ensuremath{\Proj\Tw}}
\newcommand{\Tc}{\ensuremath{\mathscr{T}}}
\newcommand{\PTc}{\ensuremath{\Proj\mathscr{T}}}

\renewcommand{\bar}[1]{\overline{#1}}
\newcommand{\inv}[1]{{#1}^{-1}}
\newcommand{\Lie}{\pounds}

\newcommand{\smallborders}{\bigl.\bigr\rvert}

\newcommand{\abs}[1]{\left|#1\right|}
\newcommand{\set}[1]{\left\{#1\right\}}

\newcommand{\chirp}[1]{\left\langle #1 \right\rangle}
\newcommand{\chirm}[1]{\left[ #1 \right]}

\newcommand{\partiald}[2]{\frac{\partial #1}{\partial #2}}

\newcommand{\alg}[1]{\ensuremath{\mathfrak{#1}}\xspace}

\newcommand{\appref}[1]{appendix~\ref{#1}}

\newcommand{\secref}[1]{section~\ref{#1}}

\newcommand{\bbT}{\mathbb{T}}

\newcommand{\calM}{\mathcal{M}}
\newcommand{\calN}{\mathcal{N}}
\newcommand{\calO}{\mathcal{O}}

\renewcommand{\G}{G}
\counterwithout{section}{part}

\begin{document}

\thispagestyle{empty}

\begin{flushright}\footnotesize
\texttt{TCDMATH 16-07}%
\end{flushright}
\vspace{1cm}

\begin{center}%
{\Large{\mathversion{bold}%
Conformal Higher Spin Theory and Twistor Space Actions
}\par}

\vspace{1.5cm}

\textrm{Philipp H\"ahnel and Tristan McLoughlin} \vspace{8mm} \\
\textit{%
School of Mathematics, Trinity College Dublin\\
College Green, Dublin 2, Ireland 
} \\
\texttt{\\ $\{$haehnel, tristan$\}$@maths.tcd.ie}

\par\vspace{15mm}

\textbf{Abstract} \vspace{5mm}

\begin{minipage}{12.16cm} 
\parindent=15pt
	We consider the twistor description of conformal higher spin theories 
and give twistor space actions for the self-dual sector of theories 
with spin greater than two that produce the correct flat space-time
spectrum. We identify a ghost-free subsector, analogous to the embedding 
of Einstein gravity with cosmological constant in Weyl gravity, 
which generates the unique spin-$s$ three-point anti-MHV amplitude 
consistent with Poincaré invariance and helicity constraints. 

	By including interactions between the infinite tower of higher-spin 
fields we give a geometric interpretation to the twistor equations of 
motion as the integrability condition for a holomorphic structure on 
an infinite jet bundle. Finally, we introduce anti-self-dual interaction 
terms to define a twistor action for the full conformal higher spin 
theory.
\end{minipage}

\end{center}

\newpage

\tableofcontents

\section{Introduction}
\label{sec:intro}
%

Maxwell's theory of electromagnetism, describing massless spin-one fields, is conformally invariant in four-dimensions. While this property is not shared by Einstein's theory of gravity, Weyl gravity and its supersymmetric generalisations \cite{Kaku:1978nz, deWit:1979dzm, Bergshoeff:1980is, Fradkin:1985am} provide examples of conformally invariant spin-two theories. Conformal higher spin theories (CHS) with $s>2$ have been previously studied at the quadratic level by Fradkin and Tseytlin \cite{Fradkin:1985am} in four dimensions. The spin-$s$ theory can be formulated in terms of a rank-$s$ completely symmetric, traceless field,  $\phi_{(\mu_1\dots \mu_s)}(x)$, with the gauge symmetry
 \begin{align}
	\delta \phi_{\mu_1\dots \mu_s}(x)= \partial_{(\mu_1}\epsilon_{\mu_2\dots \mu_s)}(x)-{\rm traces}
 \end{align}
where $\epsilon_{\mu_1\dots \mu_{s-1}}$ is an arbitrary rank-$(s-1)$ symmetric, traceless field, and the variation only involves the trace-free combination with the derivative. It was also shown that it is possible to write a gauge invariant quadratic action in terms of  a differential operator $P^{(\mu_1 \dots \mu_s) (\nu_1\dots \nu_s)}(\partial)$ of order-$2s$ in derivatives
 \begin{align}
 \label{eq:FT_hs_action}
	 S_s[\phi]=\int \dd{}^4 x\, \phi_{\mu_1 \dots \mu_s} P^{(\mu_1 \dots \mu_s) (\nu_1\dots \nu_s)}(\partial)\, \phi_{\nu_1\dots \nu_s}
 \end{align}
where $P^{(\mu_1 \dots \mu_s) (\nu_1\dots \nu_s)}=P^{ (\nu_1\dots \nu_s) (\mu_1 \dots \mu_s)}$, $P^{\mu_1}{}_{\mu_1}{}^{\dots}=0$ and $P^{\mu_1\dots }\partial_{\mu_1}=0$\,. This action results in higher derivative equations of motion which have $s(s+1)$ on-shell degrees of freedom. These theories were further studied first at the cubic level by Fradkin and Linetsky \cite{Fradkin:1989md, Fradkin:1990ps}
 and subsequently by Segal \cite{Segal:2002gd} who proposed a complete, non-linear interacting theory involving a single copy of each spin-$s$ field, $s=1, 2, 3 \dots$, and a massless scalar. 

Due to the presence of higher derivatives these theories fail to be unitary, and so their relevance as a starting point for a microscopic quantum theory is at best questionable. However they do play an important role in the study of conformal field theories (CFT), particularly in the context of the AdS/CFT correspondence. Here the higher-spin fields act as external sources and, after integrating out the CFT fields, the effective action, $W_{\rm eff}$, is a functional of the higher-spin fields. For example,  in the case of ${\mathcal N}=4$ super-Yang-Mills (SYM) on a curved background \cite{Liu:1998bu, Balasubramanian:2000pq} the effective action, after integrating out the SYM fields, consists of a logarithmically divergent piece $W_{\rm div}$ and a finite piece $W_{\rm fin}$. $W_{\rm div}$ is a functional of the fields forming the ${\mathcal N}=4$ conformal supergravity multiplet, collectively denoted $G$, and is exactly the action of ${\mathcal N}=4$ conformal supergravity
 \begin{align}
	W_{\rm div}[G]=\tfrac{N^2}{4(4\pi)^2} \ln \Lambda\, S_{\rm CSG}[G]~,
 \end{align} 
where $N$ is the rank of the SYM gauge group, and $\Lambda$ is the UV cutoff. In the limit where the ${\mathcal N}=4$ SYM is taken to be free there are infinitely many conserved traceless bilinear currents which can be coupled to conformal higher-spin fields, collectively denoted $\phi$. Expanding the resulting divergent part of the effective action, $W_{\rm div}[\phi]$, to quadratic order in higher-spin fields one finds, see \cite{Tseytlin:2002gz}, a sum over the free CHS actions \eqref{eq:FT_hs_action} of each spin. A related case is that of the free $O(N)$ vector model consisting of $N$ massless complex scalar fields, $\chi^i$. This model is conjectured \cite{Sezgin:2002rt, Klebanov:2002ja} to be dual to Vasiliev's higher spin theory \cite{Vasiliev:1990en, Vasiliev:1992av, Vasiliev:2003ev} on AdS space-time. One can minimally couple the free theory to an infinite set of symmetric, traceless Noether currents, $J^{\mu_1\dots \mu_s}\sim\chi_i^\ast \partial^{\mu_1} \dots \partial^{\mu_s}\chi^i$, and the corresponding effective action, which depends on the infinite tower of higher-spin source fields, acts as the generating functional for connected correlation functions of the currents
 \begin{align}
	W[\phi]=N \log {\rm det} (-\partial^2+\sum_{s} \phi_{\mu_1 \dots \mu_s} J^{\mu_1\dots \mu_s})~.
 \end{align}
The UV divergent part of this effective action can be taken as defining a consistent interacting theory of conformal higher-spin fields \cite{Tseytlin:2002gz, Segal:2002gd, Bekaert:2010ky}. 

It is interesting to try to find alternative formulations of these CHS theories that might illuminate some underlying structures or which are simply more convenient for calculations. Given the deep connections between twistors, the conformal group and conformal geometry%
	\footnote{%
		There are several excellent introductory textbooks to twistors, e.g. \cite{penrose1988spinors, huggett1994introduction}, and reviews e.g. \cite{woodhouse1985real,Adamo:2013cra,Wolf:2010av}; we will mostly use the notations of \cite{Adamo:2013cra}.%
	}%
, it natural to ask if there is a twistor description of conformal higher spin theories. Starting from the work of Witten \cite{Witten:2003nn} on twistor string theory the cases of spin-one, Yang-Mills, and spin-two, Weyl gravity, have been well-studied. Of particular relevance to our considerations are the twistor space actions for these theories, which will provide a model for the higher-spin case. Self-dual ${\cal N}=4$  super-Yang-Mills theory \cite{Siegel:1992xp, Chalmers:1996rq} was reformulated by Witten \cite{Witten:2003nn} as a holomorphic Chern-Simons theory on the super-twistor space $\mathbb{CP}^{3|4}$. Focusing on the non-supersymmetric gauge fields, this involves $(0,1)$-forms $A$ and $G$ on $\mathbb{CP}^{3}$, taking values in the Lie algebra of GL$(N, \mathbb{C})$ and which are respectively homogeneous of degree zero and $-4$ in the bosonic $\mathbb{CP}^3$ coordinates.

The action is 
 \begin{align}
	S=\int_{\mathbb{CP}^{3}} \Omega\wedge {\rm Tr}(G\wedge(\bar \partial A+A\wedge A))~,
 \end{align}
where $\Omega$ is a holomorphic $(3,0)$-form of degree four, and $\bar \partial$ is the Dolbeault operator on ${\mathbb{CP}^{3}}$. The extension of the twistor space action to the full theory was studied by Mason in \cite{Mason:2005zm} and found by Boels, Mason and Skinner in \cite{Boels:2006ir}. Based on considerations from twistor string theory, twistor actions for the self-dual sector of conformal gravity were proposed by Berkovits and Witten in \cite{Berkovits:2004jj}, and the extension to the full theory was also proposed by Mason in \cite{Mason:2005zm} and further studied by Adamo and Mason \cite{Adamo:2012nn, Adamo:2013tja}. A related twistor action for Einstein gravity was studied in \cite{Mason:2008jy}.

These actions are motivated in large part by the non-linear graviton construction of Penrose \cite{penrose1976nonlinear}, which provides a means to identify curved twistor spaces, $\PTc$, with self-dual space-times, that is those for which the anti-self-dual Weyl spinor vanishes $\Psi_{ABCD}=0$\,. Curved twistor spaces can be thought of as locally the same as flat twistor space $\PT={\mathbb{CP}^{3}}$ but with the complex structure deformed. A useful way to describe the deformed complex structure  \cite{eastwood1982edth, Law2001} is to modify the Dolbeault operator $\bar\partial$. Given homogeneous coordinates $Z^\alpha$, $\alpha=0,1,2,3$, on some patch, we can define a background, undeformed complex structure by the operators
 \begin{align}
	\partial = \dd Z^\alpha \frac{\partial}{\partial Z^\alpha}~, 
	~~~
	\bar{\partial} = \dd \bar Z^{\alpha'} \frac{\partial}{\partial \bar{Z}^{\alpha'}}~,
 \end{align}
which naturally defines the  splitting of the complex tangent space into ${(1,0)}$ and $(0,1)$ parts. We can deform the complex structure by adding elements of $T^{(1,0)}$ to $T^{(0,1)}$, that is we define a new, deformed Dolbeault operator 
 \begin{align}
	\bar\partial_f = \bar\partial + f ~,
 \end{align}
where $f$ is a $(1,0)$-vector valued $(0,1)$-form, $f= f^\alpha_{\alpha'}\dd \bar Z^{\alpha'} \otimes  \partiald{}{Z^\alpha}$\,.

The integrability condition for this new complex structure is the Kodaira-Spencer equation: 
 \begin{align}
	\bar\partial_f^{\,2} \equiv N^\alpha\partial_\alpha=(\bar\partial  f^\alpha+f^\beta\wedge \partial_\beta f^\alpha)\partial_\alpha=0~.
 \end{align}
It is often convenient to think of our fields as living on the non-projective twistor space, $\Tc$, in which case we must restrict to deformations which preserve the Euler vector $E=Z^\alpha \partial_\alpha$, in the sense that $\Lie_f E=-\Lie_E f=0$\,, or equivalently, $f^\alpha$ is of homogeneity degree one in $Z^\alpha$ and we must make the identification of the vector field $f$ under shifts proportional to the Euler vector field 
 \begin{align}
\label{eq:fgauge}
	f \rightarrow f + \Lambda E~,
 \end{align}
where $\Lambda$ is a $(0,1)$-form of homogeneity degree zero in $Z^\alpha$. We can fix this gauge invariance by demanding that the deformation preserves the volume form $\dd\Omega$ on $\Tc$, which imposes $\partial_\alpha f^\alpha=0$\,.

It is possible to write an action for which the corresponding equations of motion imply the integrability of the complex structure \cite{Berkovits:2004jj}. This can be done in a coordinate independent fashion \cite{Mason:2005zm} or by introducing a particular background and using explicit coordinates \cite{Adamo:2013tja}. The action functional on twistor space is given in terms of a Lagrange multiplier field $g \in \Omega^{3,0}(\PTc, \Omega^{1,1}(\PTc))$\,, imposing the constraint corresponding to integrability of the deformed complex structure.  

Introducing an appropriate basis (or working in abstract index notation)
such that the components fields are $f^{\alpha}$ and $g^\Omega_{\alpha}=g_{\alpha}\wedge \Omega$, where $\Omega$ is the holomorphic volume form, we can write the action as 
 \begin{align}
\label{eq: CG fg action}
	S_{\rm\, s.d.}[f, g] & = \int_{\PTc}  g^\Omega_\alpha \wedge N^\alpha \, .
 \end{align} 
The equations of motion following from the action are the integrability conditions for the deformed Dolbeault operator
and 
 \begin{align}
\bar\partial_f g^\Omega_\alpha	\equiv \bar\partial g^\Omega_\alpha +g^\Omega_\beta\wedge \partial_\alpha f^\beta +\partial_\beta (g^\Omega_\alpha\wedge f^\beta) = 0~.
 \end{align}
As the field $g$ must be defined on $\PTc$, we additionally have the constraint $\iota(E)g=0$\,, and with this constraint the action has the gauge invariance \eqref{eq:fgauge}.

In general the field $g$ describes an anti-self-dual excitation moving in a self-dual background. Treating the fields as small deformations of the undeformed twistor space, we can focus on the linearised theory. We now have $f^\alpha\in \Omega^{0,1}(\PT, {\cal O}(1))$\,, and 
$g_\alpha\in \Omega^{0,1}(\PT,{\cal O}(-5))$ i.e. a $(0,1)$-form of homogeneity $-5$\,. The equations of motion reduce to
 \begin{align}
	\bar \partial f^\alpha=0
	~~~{\rm and}~~~
	\bar \partial g_\alpha=0~.
 \end{align}
Additionally, the gauge invariance of the action demonstrated in \cite{Mason:2005zm} becomes $g_\alpha\rightarrow g_\alpha +\bar \partial \chi$ where $\chi$ is a section of ${\cal O}(-4)$. Consequently, on-shell, we can interpret $g_\alpha$ as defining an element in the $\bar \partial$-cohomology group $H^{0,1}(\PT, {\cal O}(-5))$, and so by the Penrose transform $g_\alpha$ corresponds to a helicity $-2$ particle. To include the self-interactions of the anti-self-dual fields and to consider amplitudes with more than a single helicity $-2$ particle one must include additional terms in the action \cite{Mason:2005zm}.

It is this construction that we wish to generalise to higher spins by considering deformations corresponding to higher-rank symmetric tensors. Starting with the spin-three case we propose Maurer-Cartan-like equations of motion for the deformations, and by introducing an appropriate Lagrange multiplier field we define a twistor space action describing the self-dual sector. The twistor fields can be defined for a general curved twistor space and so define higher-spin fields in an arbitrary self-dual geometry. However to understand the space-time interpretation we focus on the flat twistor space case, in particular showing that the twistor fields give rise, via the Penrose transform, to space-time fields satisfying the zero-rest-mass equations for a spin-three field. To make the identification between twistor fields and space-time fields clearer we construct the space-time action for the self-dual sector of the higher spin theory at the quadratic level. Finally we demonstrate that, after accounting for the gauge freedom, the on-shell spectrum matches with that of \cite{Fradkin:1985am}. 

As in the spin-two case, the on-shell representation of the Poincar\'e algebra is not diagonalisable which is a manifestation of the failure of the theory to be unitary. Maldacena \cite{Maldacena:2011mk} has argued that conformal gravity with appropriate boundary conditions is classically equivalent to Einstein gravity with a non-zero cosmological constant, $\Lambda\neq 0$. This implies \cite{Adamo:2012nn} that Einstein gravity amplitudes can be calculated in conformal gravity by restricting to the asymptotic states of Einstein gravity and accounting for the appropriate powers of the cosmological constant. Adamo and Mason \cite{Adamo:2012xe, Adamo:2013tja} studied supergravity scattering amplitudes by performing such a truncation of conformal gravity in the twistor space description and were able to show that the resulting determinant formulae was directly related to Hodges' formula \cite{Hodges:2012ym}. We perform an analogous truncation on the higher-spin action to identify a ``unitary" sub-sector of CHS theory. At the quadratic level this sector has the usual Fronsdal \cite{Fronsdal:1978rb, Fang:1978wz, deWit:1979sib} spectrum of massless higher spins and we show that the ${\overline{\rm MHV}}$ three-point amplitude agrees with the constraints from Poincar\'e invariance \cite{Benincasa:2007xk}.  

In order to interpret the higher-spin deformation in a geometric sense we must include an infinite number of interacting higher-spin fields. This is because once we go beyond the spin-two case the Maurer-Cartan equations for a single spin can no longer be interpreted as the integrability condition for a holomorphic structure on a vector bundle. This can be rectified by including an infinite tower of interacting spins and by interpreting the deformations as acting on the corresponding infinite jet bundle of the space of symmetric products of the (co)tangent bundle. We discuss this formulation in \secref{sec:cons_ints}. Finally in \secref{sec:asd_int_act} we describe additional terms to be added to describe the interactions of the anti-self-dual fields extending the action to the full higher spin theory.

\section{Spin-Three Fields }

\subsection{Spin-Three Fields in Twistor Theory}

We start our study of the higher spin theory with the case of spin-three fields. We again pick some background twistor space $\Tc$ and its projective version $\PTc$. We will often take $\Tc$ to be flat twistor space $\Tw$, however, if we wish to consider higher-spin fields on a general self-dual background, it will be the corresponding curved twistor space. We give a brief review of some salient aspects of curved twistor spaces and a description of our notations in \appref{sec:CurvedTwistorSpace}. We will take $Z^\alpha$ to be homogeneous coordinates on the background twistor space. The complex deformations corresponding to the spin-two case discussed in the introduction are $(0,1)$-forms taking values in the $(1,0)$ part of the tangent space, that is they are elements of $\Omega^{0,1}(\PTc, T^{1,0}(\PTc))$. We now consider deformations corresponding to $(0,1)$-forms taking values in symmetric products of the tangent space, that is they are elements of $\Omega^{0,1}(\PTc, \mathrm{Sym}^2(T^{1,0}(\PTc)))$, 
 \begin{align}
	f^{(2)} = f^{\alpha\beta}\partial_\alpha \otimes \partial_\beta ~ .
 \end{align}
This twistor field will be interpreted in space-time as a massless spin-three field and we will refer it as such. As it is in fact defined on projective space, rather than the full twistor space, we must make the identification analogous to \eqref{eq:fgauge} for the field $f$
 \begin{align}
\label{eq:f3gauge}
	f^{(2)}\rightarrow f^{(2)}+E\otimes \Lambda+\Lambda \otimes E~,
 \end{align}
where $\Lambda \in \Omega^{0,1}(\PTc, T^{1,0}(\PTc))$ is now vector-valued, and $E$ is again the background Euler vector field. More particularly, we consider an operator 
 \begin{align}
	\bar\partial_f :\Omega^{p,q}(\PTc)\rightarrow \Omega^{p,q+1}(\PTc)
 \end{align}
acting on forms $k\in \Omega^{0,1}(\PTc)$ as
 \begin{align}
	k\mapsto \bar \partial k+f^{\alpha\beta}\wedge \partial_\alpha\partial_\beta k \in  \Omega^{0,2}(\PTc) ~ .
 \end{align}
Such an operator can be naturally viewed as a deformation of the Dolbeault operator, however, due to the presence of higher derivatives, it is not a derivation. 

We will impose the condition that the deformation $f$ satisfies the equation
 \begin{align}
	N^{(2)} \equiv \left(\bar\partial f^{\alpha\beta} + f^{\gamma\delta}\wedge \partial_\gamma \partial_\delta f^{\alpha\beta}\right)\partial_\alpha \partial_\beta=0 ~ .
 \end{align}
This condition does not imply $\bar \partial_f^{\,2} = 0$\,; however, as we will see later, this failure can be compensated by including an infinite tower of higher-spin fields. The necessity of such an infinite tower is unsurprising given known results in space-time formulations. However, we will postpone a discussion of this, and here only consider the self-interactions of the spin-three fields. 

As for the spin-two case, to define an action functional we introduce a corresponding Lagrange multiplier field $g$ which takes values in the dual space, i.e. 
 \begin{align}
	g^{(2)}\in \Omega^{0,1}(\PTc, {\rm Sym}^2(T^{\ast,1,0}(\PTc)))\otimes \Omega~.
 \end{align}
As before, $\Omega$ is a section of $\Omega^{3,0}(\PT)\otimes {\cal O}(4)$\,, and we thus have
 \begin{align}
	g^{(2)}
  = (g_{\alpha\beta}\wedge \Omega) \otimes  \dd Z^\alpha \otimes \dd Z^\beta ~,
 \end{align}
where $g_{\alpha\beta} = g_{(\alpha\beta)} \in \Omega^{0,1}(\PTc, \calO(-6))$\,. To ensure the appropriate gauge invariance we impose the constraint $g_{\alpha\beta}Z^\alpha = 0$\,.
The twistor space action for the self-dual sector is proposed to be the obvious analogue of the spin-two case
 \begin{align}
	S_{{\rm\, s.d.}}\left[f^{(2)}, g^{(2)}\right] & = \int_{\PTc} \Omega \wedge g_{\alpha\beta} \wedge \left(\bar\partial f^{\alpha\beta} + f^{\gamma\delta}\wedge \partial_\gamma \partial_\delta f^{\alpha\beta}\right)
	~,
 \end{align}
from which follows that $N^{(2),\alpha\beta}=0$ as required, and also 
 \begin{align}
	\bar \partial g^\Omega_{\alpha\beta}
	 + g^\Omega_{\gamma\delta}\wedge \partial_\alpha \partial_\beta f^{\gamma\delta}
	 - \partial_\gamma \partial_\delta(g^\Omega_{\alpha\beta} \wedge f^{\gamma \delta}) = 0~,
 \end{align}
where $g^\Omega_{\gamma \delta} = g_{\gamma\delta}\wedge \Omega$\,. In local holomorphic coordinates, $Z^\alpha$, we will often write $\Omega=\DD[3] Z$\,, where $\DD[3] Z$ is the usual weight four holomorphic volume form. 

We will next be interested in understanding the spectrum of this action corresponding to its quadratic approximation about a given background. We thus focus on the linearised equations of motion, which are straightforwardly given by
 \begin{align}
	\bar \partial f^{\alpha\beta} = 0
	~~~{\rm and}~~~
	\bar \partial g_{\alpha\beta} = 0~.
 \end{align}
Also at the linearised level, the action has the additional gauge invariance $g_{\alpha\beta}\rightarrow g_{\alpha\beta}+\bar \partial \chi_{\alpha\beta}$\,. Consequently, we can think of $g_{\alpha\beta}$ as defining an element in the Dolbeault cohomology $H^{0,1}(\PTc, {\cal O}(-6))$, satisfying $g_{\alpha\beta}Z^\alpha=0$\,.

\subsection{Space-Time Interpretation}

In the twistor description of conformal gravity we start with two tensor-valued forms, $f^{\alpha}(Z)$ and $g_{\alpha}(Z)$, defined on $\PTc$ of homogeneity degree $1$ and $-5$ respectively. These fields correspond, via the Penrose transform, to the space-time anti-self-dual Weyl spinor $\Psi_{ABCD}$, which vanishes for self-dual backgrounds, and the Lagrange multiplier field  $\Gamma_{ABCD}$, which satisfies a second order equation of motion \cite{Mason:2005zm}. In the  higher-rank spin-three generalisation above, we have the fields $ f^{\alpha_1\alpha_2 }(Z)$, of homogeneity degree $2$, and $g_{\alpha_1\alpha_2}(Z)$, of homogeneity degree $-6$. The zero-rest-mass equations on space-time can be found by using the Penrose transform for such tensors as described in \cite{mason1990local}, or see \cite{Adamo:2013cra} for a recent review. 

In order to consider the Penrose transform of such tensors it is necessary to choose a specific frame. We will focus on the case of flat space-time $M$, and so consider the twistor fields as deformations of flat twistor space, however it is straightforward to generalise to deformations of an arbitrary curved twistor space and so find the spin-three fields. We consider the bundle
 \begin{align}
	{\mathbb T}^{\alpha}{}' \rightarrow \mathbb{PT}(M)
 \end{align}
whose local holomorphic sections are represented by vector fields $T$ on $\mathbb{T}(M)$, satisfying
 \begin{align}
	[E, T]=-T
	~~~{\rm and}~~~ 
	[ \overline{E}, T]=0 ~,
 \end{align}
where $E$ is the Euler vector field. This can be considered as the pull-back of the local twistor bundle $ {\mathbb T}^{\alpha}\rightarrow M$ with fibre coordinates $(\lambda_A, \mu^{A'})$. Choosing a basis on ${\mathbb T}^{\alpha}{}'$, which we denote $\delta^\alpha_{\underline \alpha}$, $\alpha=0,1,2,3$, such that 
 \begin{align}
	\delta^\alpha_{\underline \alpha}\frac{\partial}{\partial Z^\alpha}
 \end{align}
is a global holomorphic frame, the components of the tensors corresponding to the spin-two and  spin-three fields with respect to the frame (or its dual) are 
 \begin{align}
	f^{\underline \alpha}\,, ~
	g_{\underline \alpha}\,, ~
	f^{{\underline \alpha_1}{\underline \alpha_2}}\,, ~
	g_{{\underline \alpha_1}{\underline \alpha_2}}\,, ~
	{\rm etc}~.
 \end{align}
For each component we can perform the Penrose transform. Considering the spin-three case we have a $(0,1)$-form-valued tensor $g_{{\underline \alpha_1}{\underline \alpha_2}}$ whose components have homogeneity $-6$. This corresponds to the case in \cite{mason1990local} of a tensor with homogeneity $m-2$\,, with $m<0$, and so in the Penrose transform when performing the integration over the projective complex line $X\simeq \CP^1$, corresponding to the space-time point $x$,
 \begin{align}
	\G_{({\underline \alpha_1}{\underline \alpha_2})(B_1\dots B_4)} 
		= \frac{1}{2\pi i}\int_X \left. g_{{\underline \alpha_1}{\underline \alpha_2}} \lambda_{B_1}\dots \lambda_{B_4}\right|_X \wedge \DD \lambda ~ ,
 \end{align}
we include four factors of  $\lambda_A$, the homogeneous coordinates on $X$, to compensate for the weight of $g_{{\underline \alpha_1}{\underline \alpha_2}}$ and holomorphic measure on $X$ $\DD \lambda = \chirp{\lambda \dd\lambda}$\,. The resulting space-time field $\G_{({\underline \alpha_1}{\underline \alpha_2})(B_1\dots B_4)}$ satisfies the zero-rest-mass equation
 \begin{align}
	\nabla^{B_1'B_1} \G_{({\underline \alpha_1}{\underline \alpha_2})(B_1\dots B_4)} = 0 ~ .
 \end{align}
Due to the specific choice of frame we can think of the twistor space tensor indices as local twistor indices. In particular the covariant derivative acts on the tensors with twistor indices by the local twistor connection, which for flat space gives
\begin{gather}
\label{eq:zrmeq}
 \nabla^{B_1'B_1} \G_{A_1' A_2' B_1\dots B_4} = 0 ~,\\
	 \nabla^{B_1'B_1} \G^{A_1 }{}_{A_2' B_1\dots B_4} - i \epsilon^{B_1 A_1} \G^{B_1'}{}_{A_2'B_1\dots B_4}  = 0 ~,\nn\\
	 \nabla^{B_1'B_1} \G^{A_2 }{}_{A_1' B_1\dots B_4} - i \epsilon^{B_1 A_2} \G^{B_1'}{}_{A_1'B_1\dots B_4} = 0 ~,\nn\\
	 \nabla^{B_1'B_1} \G^{A_1 A_2}{}_{B_1\dots B_4}
		 - i \epsilon^{B_1 A_1}\G^{B_1'A_2}{}_{B_1\dots B_4}
		 - i \epsilon^{B_1 A_2}\G^{B_1'A_1 }{}_{B_1\dots B_4}  = 0 ~.\nn
\end{gather}
We can use these to determine all the components of $\G_{({\underline \alpha_1}{\underline \alpha_2})(B_1\dots B_4)}$ in terms of the derivatives of the fields $\G^{ A_1 A_2}{}_{B_1 \dots B_4}$.
Furthermore we must impose the twistor space constraint on the Lagrange multiplier field $Z^{\underline  \alpha_1} g_{{\underline \alpha_1} {\underline \alpha_2}}=0$\,, which corresponds to acting with a helicity lowering operator and imposes the space-time condition 
 \begin{align}
	\G^{A_1}{}{}_{{\underline \alpha_2} A_1\dots B_4} = 0~,
 \end{align}
namely that the field $\G_{A_1 A_2 B_1\dots B_4}$ is symmetric in all its indices. Using this and \eqref{eq:zrmeq} we find that the anti-self-dual field satisfies the third-order equations 
 \begin{align}
	\nabla^{A_1'A_1} \nabla^{A_2'A_2} \nabla^{A_3'A_3}\G_{A_1A_2A_3A_4A_5A_6} = 0~,
 \end{align}
which is to say that $\G_{A_1A_2A_3A_4A_5A_6}$ satisfies the zero-rest-mass equation for a spin-three field. 

As the tensor field $f^{{\underline \alpha_1}{\underline \alpha_2}}$ has homogeneity $m-2=2$, $m>0$, in a general background it transforms into a potential. This is exactly analogous to the conformal gravity spin-two case where the field $f^\alpha$ of homogeneity one describes the chiral Weyl spinor $\Psi_{ABCD}$\,, see \cite{mason1987relationship}. In the linearised spin-two theory the relationship with the metric fluctuation, $h_{ABC'D'}$, is given by
 \begin{align}
	\Psi_{ABCD}=\nabla^{C'}_{(C}\nabla^{D'}_{D} h_{AB)C'D'}~.
 \end{align}
This arises by taking the Penrose transform of $f^\alpha$ which gives the potential $\Sigma^{\underline \alpha}{}_{B_1B_2 C'}$, which can be decomposed as 
 \begin{align}
	\Sigma^{\underline \alpha}{}_{B_1B_2 C'} = \left(\tilde \Sigma_{(AB_1B_2) C'}, i h_{B_1 B_2}{}^{A'}_{C'}\right) ~,
 \end{align}
where the symmetry in the unprimed indices follows from using the gauge symmetry of $f^\alpha\rightarrow f^\alpha+Z^\alpha \Lambda$\,. The condition that the solution be self-dual, that is $\Psi_{ABCD}=0$, follows from the condition on the potential 
 \begin{align}
	\nabla^{C'}{}_{(C}\Sigma^{\underline \alpha}{}_{B_1B_2)C'}=0~,
 \end{align}
and by using the local twistor connection as above. In the spin-three case we will have a higher-rank potential
 \begin{align}
	\Sigma^{{\underline \alpha_1}{\underline \alpha_2}}{}_{B_1B_2B_3 C'}~,
 \end{align}
which will give rise to fields analogous to the metric fluctuations $\phi_{A_1A_2A_3}{}^{B_1'B_2'B_3'}$, which satisfy a condition $\Gamma_{A_1A_2A_3B_1B_2B_3}=0$, where $\Gamma_{A_1A_2A_3B_1B_2B_3}$ is the spin-three analogue of the Weyl spinor, $\Gamma_{A_1A_2A_3B_1B_2B_3}$, with higher numbers of derivatives
 \begin{align}	\Gamma_{A_1A_2A_3B_1B_2B_3}=\nabla^{A_1'}_{(A_1}\nabla^{A_2'}_{A_1}\nabla^{A_3'}_{A_3} \phi_{B_1B_2B_3)A_1'A_2'A_3'} ~.
 \end{align}
It is interesting to relate this description of a self-dual sector of the conformal spin-three theory, even at just the quadratic level, with the usual formulation of conformal higher spin theory, both to make the relation clearer and to let us use this connection to motivate an action for the full theory.

\paragraph{A linearised space-time action:}

As we have seen, the higher spin twistor theory gives a space-time analogue of the anti-self-dual Weyl spinor $\Gamma_{A_1A_2A_3B_1B_2B_3}$. Motivated by this we consider the combination of self-dual and anti-self-dual spinor fields $\Gamma_{A_1A_2A_3B_1B_2B_3} $ and ${\tilde \Gamma}_{A_1'A_2'A_3'B_1'B_2'B_3'}$ and we define the field strength 
 \begin{align}
\label{eq:s3fs}
	C_{a_1 b_1 a_2 b_2 a_3 b_3 }
		= \epsilon_{A_1'B_1'}\epsilon_{A_2'B_2'}\epsilon_{A_3'B_3'} \Gamma_{A_1A_2A_3B_1B_2B_3}
			+ \epsilon_{A_1 B_1}\epsilon_{A_2 B_2}\epsilon_{A_3 B_3} {\tilde \Gamma}_{A_1'A_2'A_3'B_1'B_2'B_3'} ~.
 \end{align}
This tensor is anti-symmetric in each pair $(a_i, b_i)$ for $i=1, 2, 3$ and symmetric between pairs 
$(a_i, b_i)\leftrightarrow (a_j, b_j)$ for $i\neq j$. Moreover, if we contract over a pair of $a$ or $b$ indices we find zero, for example
 \begin{align}
	C^{a_1}{}_{ b_1 a_1 b_2 a_3 b_3 }=0~.
 \end{align}
This is essentially the spin-three example of the higher-spin curvatures introduced by Weinberg \cite{Weinberg:1964ew}, which are equivalent to those introduced by de Wit and Freedman \cite{deWit:1979sib}; for the relation between these formulations and further relevant discussion see \cite{Marnelius:2008er, Vasiliev:2009ck}. 
 
We can reformulate this in a notation closer to that used by Fradkin and Tseytlin in their discussion of quadratic higher spin theory \cite{Fradkin:1985am} by introducing a potential for this field strength, $\phi_{b_1b_2b_3}$, which is symmetric and pairwise traceless in its indices. The field strength
 \begin{align}
	C_{a_1 b_1 a_2 b_2 a_3 b_3 } = \partial_{a_1}\partial_{a_2}\partial_{a_3}\phi_{b_1b_2b_3} - \partial_{b_1}\partial_{a_2}\partial_{a_3 }\phi_{a_1b_2b_3} \pm {\rm permutations}
 \end{align}
is found by anti-symmetrising on pairs of indices $a_i$, $b_i$. We can naturally form a Lagrangian density quadratic in the field strength
 \begin{align}
	{\cal L}_{{\rm spin}-3} & =\frac{1}{64} C^{a_1 b_1 a_2 b_2 a_3 b_3 }C_{a_1 b_1 a_2 b_2 a_3 b_3 }~.
	\end{align}
	Rewriting this in terms of the potential and neglecting total derivative terms
we have
\begin{align}
	&\frac{1}{8} \phi^{c_1 c_2 c_3} P_{c_1c_2c_3}^{b_1b_2b_3} \left( \vphantom{\frac12}
				\Box^3 \delta_{b_1}^{a_1}\delta_{b_2}^{a_2}\delta_{b_3}^{a_3} - 3 \Box^2 \partial^{a_1}\partial_{b_1} \delta_{b_2}^{a_2}\delta_{b_3}^{a_3} + 3\Box \partial^{a_1}\partial^{a_2}\partial_{b_1}\partial_{b_2}\delta^{a_3}_{b_3} \right.\nn\\
				& \kern+80pt \left. \vphantom{\frac12} - \partial^{a_1}\partial^{a_2}\partial^{a_3}\partial_{b_1}\partial_{b_2}\partial_{b_3}\right) P_{a_1a_2a_3}^{d_1d_2d_3} \phi_{d_1 d_2 d_3}
	~,\nn
\end{align}
where $P^{a_1 a_2 a_3}_{b_1b_2 b_3}$ projects onto symmetric, pairwise traceless tensors. Thus we can write the Lagrangian 
\begin{align}
	{\cal L}_{{\rm spin}-3} & = \frac{1}{2} \phi^{b_1 b_2b_3}D^{a_1 a_2 a_3}_{b_1b_2 b_3}\phi_{a_1a_2a_3} 
 \end{align}
in terms of a kinetic operator $D^{a_1 a_2 a_3}_{b_1b_2 b_3}$ which is symmetric, pairwise traceless and transverse, that is satisfying
 \begin{align}
	D^{a_1 a_2 a_3}_{b_1 b_2 b_3}\partial_{a_1} = 0~.
 \end{align}
The generalisation of this construction to higher-spin fields is immediate, and the resulting Lagrangian has the form of the conformal higher spin theory described in \cite{Fradkin:1985am}. 

In terms of the spinor fields the quadratic action for spin-three fields is 
 \begin{align}
	S_{\rm spin-3} & = \frac{1}{ \lambda}\int \dd{}^4 x\, {\cal L}_{{\rm spin}-3} 
		= \frac{1}{8 \lambda}\int \dd{}^4 x\, \left(\Gamma^{A_1\dots B_3}\Gamma_{A_1 \dots B_3}
			+ {\tilde \Gamma}^{A_1' \dots B_3'}{\tilde \Gamma}_{A_1' \dots B_3'} \right) ~ ,
 \end{align}
 where we have introduced a dimensionless parameter, $\lambda$.

\paragraph{A self-dual sector:}

The action for conformal gravity can be expressed in terms of $\Psi_{ABCD}$ and its dual $\tilde \Psi_{A'B'C'D'}$ as 
 \begin{align}
	S_{\rm spin-2}=\frac{1}{\lambda }\int \sqrt{g}\dd{}^4x\, ( \Psi^{ABCD} \Psi_{ABCD}+{\tilde \Psi}^{A'B'C'D'} {\tilde \Psi}_{A'B'C'D'})~.
 \end{align}
By adding a topological term this action can be written as \cite{Berkovits:2004jj}
 \begin{align}
	S_{\rm spin-2}&=\frac{1}{4 \lambda }\int \sqrt{g}\dd{}^4x\,  \Psi^{ABCD} \Psi_{ABCD}~,\\
	 &=\int \sqrt{g}\dd{}^4x\,  G^{ABCD} \Psi_{ABCD}
	 - \lambda \int \sqrt{g}\dd{}^4x\,  G^{ABCD} G_{ABCD}~.
 \end{align}
The $\lambda \to 0$ limit describes the self-dual sector of the theory, while the second term describes the self-interactions of the anti-self-dual modes. It is this form of action which is most closely connected with the twistor space description. To make contact with the twistor description of the higher spin theory, we wish to define an analogous self-dual sector.

Given the symmetries of the field strength ${C}_{a_1 b_1 a_2 b_2 a_3 b_3 }$, we can naturally define a dual field strength
 \begin{align}
	{\cal  C}_{a_1 b_1 a_2 b_2 a_3 b_3 }\equiv\ast C_{a_1 b_1 a_2 b_2 a_3 b_3 } = \epsilon_{a_1 b_1 c_1 d_1} C^{c_1 d_1}{}_{ a_2 b_2 a_3 b_3 }~.
 \end{align}
In terms of spinor quantities the Levi-Civita tensor is
 \begin{align}
	\epsilon_{a  b c d}=i \epsilon_{AC}\epsilon_{BD}\epsilon_{A'D'} - i \epsilon_{AD}\epsilon_{BC}\epsilon_{A'C'}\epsilon_{B'D'}
 \end{align}
such that, by construction, the anti-self-dual part is
 \begin{align}
	{}^-C_{a_1 b_1 a_2 b_2 a_3 b_3} 
		& = \tfrac{1}{2}(C_{a_1 b_1 a_2 b_2 a_3 b_3}+ i \ast C_{a_1 b_1 a_2 b_2 a_3 b_3})\nn\\
		& = \epsilon_{A_1'B_1'}\epsilon_{A_2' B_2'}\epsilon_{A_3' B_3'} \Gamma_{A_1 A_2 A_3 B_1 B_2 B_3} ~,\nn\\
\intertext{and the self-dual part is}
	{}^+C_{a_1 b_1 a_2 b_2 a_3 b_3} 
		& = \tfrac{1}{2}(C_{a_1 b_1 a_2 b_2 a_3 b_3}- i \ast C_{a_1 b_1 a_2 b_2 a_3 b_3}) \\
		& =\epsilon_{A_1B_1}\epsilon_{A_2 B_2}\epsilon_{A_3 B_3}{\tilde \Gamma}_{A_1' A_2' A_3' B_1' B_2' B_3'}~.\nn 
 \end{align}
It is straightforward to show that the term
 \begin{align}
	i \ast C^{a_1 b_1 a_2 b_2 a_3 b_3}C_{a_1 b_1 a_2 b_2 a_3 b_3} 
		& =  \Gamma^{A_1 \dots B_3}\Gamma_{A_1 \dots B_3} - {\tilde \Gamma}^{A_1' \dots B_3'}{\tilde \Gamma}_{A_1' \dots B_3'}
 \end{align}
is a total derivative and so does not affect any perturbative calculations. Following the construction of the Chalmers-Siegel action for Yang-Mills \cite{Chalmers:1996rq, Chalmers:1997sg} and its analogue for conformal gravity \cite{Berkovits:2004jj}, though of course here we are working only at the linearised level, we can add this term to the action so that, up to boundary terms, we find 
 \begin{align}
\label{eq:spin3action}
	S_{\rm spin-3} &= \frac{1}{4\lambda}\int \dd{}^4 x\, \Gamma^{A_1 \dots B_3}\Gamma_{A_1\dots B_3} \nn\\
		&= \int \dd{}^4 x \, G^{A_1 \dots B_3}\Gamma_{A_1 \dots B_3} 
			- \lambda \int \dd{}^4x \, G^{A_1 \dots B_3}G_{A_1 \dots B_3}~,
 \end{align}
where in the last line we have introduced the anti-self-dual Lagrange multiplier field $G^{A_1 \dots B_3}$, which is symmetric in all indices. This action gives the equations of motion 
 \begin{align}
	\nabla^{A_1A_1'} \nabla^{A_1A_1'}\nabla^{A_1A_1'} G^{A_1 \dots B_3}=0~,
	~~~ \Gamma^{A_1 \dots B_3} = 2 \lambda G^{A_1 \dots B_3}~.
 \end{align}
If we set the parameter $\lambda$ to zero we find the self-dual theory described by the twistor fields above, with the $g_{\alpha_1 \alpha_2}$ corresponding to the space-time Lagrange multiplier field by the Penrose transform. This action will also suggest, much as in the case of conformal gravity, how to extend the twistor action to the full theory beyond the self-dual sector.

\subsection{Minkowski Space-Time Spectrum}
\label{ssec:Mspec}

Given the matching of the equations of motion, it should be unsurprising that the counting of the on-shell degrees of freedom in both the twistor and space-time descriptions of CHS theory also agrees. Nonetheless it provides a useful check and provides further insight into the twistor description, particularly the appearance of the ghost degrees of freedom, which result in the theory failing to be unitary. To this end we wish to understand the flat-space spectrum, corresponding to the twistor fields $f^{\alpha\beta}(Z)$ of homogeneity $n=2$ and $g_{\alpha\beta}(Z)$ of homogeneity $n=-6$, while taking into account the gauge invariance and constraint, respectively
 \begin{align}
\label{eq:proj_const}
	& f^{\alpha \beta} \rightarrow  f^{\alpha \beta}+Z^{(\alpha}\Lambda^{\beta)}~, 
	~~~{\rm and}~~~ 
	g_{\alpha \beta} Z^\alpha =0~.
 \end{align}
In the standard application of the Penrose transform, a function of the homogeneous coordinates $Z^{\alpha}$ with homogeneity degree $n$ corresponds to a massless state of helicity $s=1+n/2$\,. In the case at hand we must further take into account the indices $\alpha$, $\beta$ etc. One way to do this is, following \cite{Berkovits:2004jj}, to form invariant functions using the flat space twistor coordinates $\lambda_{A}$ and $\mu^{A'}$, which then correspond to definite helicity states. For example there are three homogeneity four functions or $s=3$ states 
\footnote{
	As we are considering flat space we can raise and lower spinor indices using $\epsilon_{AB}$ and $\epsilon_{A'B'}$ as required.
	}
 \begin{align}
	\lambda_A \lambda_B f^{AB}~, 
	\qquad \mu^{A'}\mu^{B'}f_{A'B'}~, 
	\qquad \lambda_A\mu^{A'}f^{A}{}_{B'}~.
 \end{align}
Additionally, we may form invariants using derivatives, $\partial_A=\tfrac{\partial}{\partial \lambda^A}$ and $\partial^{A'}=\tfrac{\partial}{\partial \mu_{A'}}$\,, so there are four homogeneity two functions or $s=2$ states, 
 \begin{align}
	\lambda_A \partial_B f^{AB}~, 
	\qquad \mu^{A'}\partial^{B'}f_{A'B'}~, 
	\qquad \lambda_A\partial^{A'}f^{A}{}_{B'}~, 
	\qquad \mu^{A'}\partial_{A}f^{A}{}_{B'}~,
 \end{align}
and finally there are three $s=1$ states 
 \begin{align}
	\partial_A \partial_B f^{AB}~, 
	\qquad \partial^{A'}\partial^{B'}f_{A'B'}~, 
	\qquad \partial_A\partial^{A'}f^{A}{}_{B'}~.
 \end{align}
Hence there are ten on-shell degrees of freedom, however some of these are simply gauge and can be removed by a transformation using $\Lambda^\alpha$. In particular we can form the invariants
 \begin{align}
	\lambda_A \Lambda^A ~, 
	\qquad \mu^{A'}\Lambda_{A'}~,  
	\qquad \partial_A \Lambda^A~,
	\qquad \partial^{A'} \Lambda_{A'}
 \end{align}
and so remove four degrees of freedom. Specifically, we can use this freedom to set $\partial_\alpha f^{\alpha\beta}=0$ and so remove two states with $s=2$ and two corresponding to $s=1$. Hence we find a total of six on-shell states from the tensor field $f^{\alpha\beta}$. We can repeat this argument for $g_{\alpha\beta}$, for which we have ten invariants
$\lambda^A \lambda^B g_{AB}$, $\mu_{A'}\mu_{B'}g^{A'B'}$, $\lambda^A\mu_{A'}g_{A}{}^{B'}$, 
$\lambda^A \partial^B g_{AB}$, $\mu_{A'}\partial_{B'}g^{A'B'}$,  $\lambda^A\partial_{A'}g_{A}{}^{B'}$,  $\mu_{A'}\partial^{A}g_{A}{}^{B'}$,
$\partial^A \partial^B g_{AB}$,  $\partial_{A'}\partial_{B'}g^{A'B'}$,  $\partial^A\partial_{A'}g_{A}{}^{B'}$,
which correspond to three states of $s=-3$, four of $s=-2$ and three of $s=-1$. Four degrees of freedom are removed by the constraint $Z^\alpha g_{\alpha\beta}=0$, of which two are $s=-2$ and two are $s=-1$. Hence in total we find twelve on-shell degrees of freedom. 
 
We could alternatively have made use of the duality between $f^{\alpha\beta}$ and $g_{\alpha\beta}$, \textit{c.f.} \cite{Berkovits:2004jj}, following from the Fourier-like transform to the dual twistor space described by coordinates $W_\alpha$:
 \begin{align}
	\tilde{g}^{\alpha\beta}(W)=\int \DD[3] Z \, f^{\alpha\beta}(Z) \exp({W\cdot Z})~.
  \end{align}
The homogeneity in $W$ of $n=-6$ for $\tilde{ g}^{\alpha\beta}$ follows immediately from the homogeneity in $Z$ of $n=2$ for $f^{\alpha\beta}$ and the weight, $4$, of the measure. Similarly, the gauge invariance of $f^{\alpha\beta}$ implies $\tilde{ g}^{\alpha\beta}\rightarrow \tilde{ g}^{\alpha\beta}+\partial^{(\alpha}\Lambda^{\beta)}$ which can be fixed by imposing the constraint \eqref{eq:proj_const}. Thus we expect to find the same number of space-time states described by ${\tilde g}^{\alpha\beta}$
as by $f^{\alpha\beta}$, namely six.

\paragraph{A unitary subsector:}

As is well known, conformal higher spin theories are not unitary. One symptom of this is the fact that the representation of certain Poincar\'e generators on on-shell states is not diagonalisable. For example, the generator of space-time translations on twistor space is given by  the vector field
 \begin{align}
	P_{AA'}=\lambda_A\frac{\partial}{\partial \mu^{A'}}~.
 \end{align}
Its action on twistor space contravariant tensors $f^{\alpha_1\dots \alpha_n}$ can be calculated straightforwardly from the Lie derivative 
 \begin{align}
	\Lie_{P_{AA'}}f^{\alpha_1\dots} = \lambda_A\frac{\partial}{\partial \mu^{A'}}f^{\alpha_1\dots} - \sum_{i=1}^n \delta^{\alpha_i}_{A'} f\indices{^{\alpha_1\dots\alpha_{i-1}}_{A}^{\alpha_{i+1}\dots\alpha_n}}~,
 \end{align}
which can be seen to be non-diagonalisable. For example, in the case of conformal gravity it was shown, \cite{Berkovits:2004jj}, that the generator $P_{AA'}$ acting on the pair $\begin{pmatrix}f_B\\ f^{B'} \end{pmatrix}$ is represented by the non-diagonalisable matrix  
 \begin{align}
	P_{AA'} =\begin{pmatrix} \lambda_A\frac{\partial}{\partial \mu^{A'}} & 0 \\ \ast & \lambda_A\frac{\partial}{\partial \mu^{A'}} \end{pmatrix}~.  
 \end{align}
That such generators are not Hermitian is one aspect of the lack of unitarity of the full theory. One can however truncate to a unitary sector by restricting to space-time fields corresponding to twistor components $f^{A'}$ and $g^A$. Exactly analogous arguments can be made for higher-rank symmetric twistor tensors where we must restrict to contravariant tensors with only primed indices $f_{A'_1\dots A'_n}$ and unprimed covariant tensors $g^{A_1\dots A_n}$. This can be phrased in an alternative manner by making use of the infinity twistor ${ I}^{\alpha\beta}$.

In a homogeneous space-time with cosmological constant $\Lambda$ we can write the infinity twistor and its dual as
 \begin{align}
\label{eq:infinity_twistor_lambda}
	{ I}_{\alpha\beta}=\begin{pmatrix} \epsilon^{AB} & 0 \\ 0 & \Lambda \epsilon_{A'B'}\end{pmatrix}~, 
	~~~ {I}^{\alpha\beta}=\begin{pmatrix} \Lambda \epsilon_{AB} & 0 \\ 0 &\epsilon^{A'B'}\end{pmatrix}~,
 \end{align}
 where $I^{\alpha\beta} I_{\beta\gamma} = -\Lambda \delta^\alpha_\gamma$\,.
A restriction to wave-functions of the form 
 \begin{align}
	 \begin{aligned}
	f^\alpha(Z) &=I^{\beta \alpha}\partial_\beta h(Z) = \left(\Lambda \epsilon_{BA} \partial^B h(Z), \epsilon^{B'A'} \partial_{B'} h(Z)\right) \\
	g_\alpha(Z) &=I_{\alpha\beta} Z^\beta \tilde h(Z) = \left(\epsilon^{AB} \lambda_B \tilde h(Z), \Lambda \epsilon_{A'B'} \mu^{B'} \tilde h(Z)\right)
	 \end{aligned}
 \end{align}
exactly corresponds to the restriction to twistor components $f^{A'}$ and $g^A$ on flat space and gives the correct extension to the case of  non-vanishing cosmological constant. This is the twistor description of the restriction of conformal gravity to Einstein gravity \cite{Adamo:2012nn, Adamo:2013tja}. By analogy, for the case of $n=2$, we may consider the truncation of the spin-three theory to fields of the form
 \begin{align}
 \label{eq:spin_3_unitary}
	 \begin{aligned}
	f^{\alpha\beta}(Z) & = I^{\alpha\gamma} I^{\beta\delta} \partial_\gamma \partial_\delta h(Z) \,, \qquad
	g_{\alpha\beta}(Z) = I_{\alpha\gamma} I_{\beta\delta} Z^\gamma Z^\delta \tilde h(Z)  \,.
	 \end{aligned}
 \end{align}
At the linearised level these fields are such that on-shell $h\in H^{0,1}(\PTc, \calO(4))$ and $\tilde h\in H^{0,1}(\PTc, \calO(-8))$, and so they describe space-times fields of helicity $\pm 3$. Thus we have at the quadratic level a massless, ghost free, spin-three theory. The obvious question is whether the twistor theory describes consistent interactions, and in order to understand the self-interactions we will consider the on-shell amplitudes, at least for three particles. However for the spin-three case this is essentially trivial. We could compute the three-point function for arbitrary asymptotic states and then restrict to the unitary sector, but it is in this case sufficient to make the truncation directly in the action. In particular, if we focus on the cubic terms and substitute the expressions \eqref{eq:spin_3_unitary} we find that the action reduces to 
 \begin{align}
	S_{\rm s.d.}^{\rm 3pt}\left[f^{(2)},g^{(2)}\right] 
			& = 2\Lambda^2 \int_{\PTc} \DD[3]Z \wedge \tilde h \wedge I^{\gamma \sigma} I^{\delta \tau} \left(\partial_\gamma \partial_\delta h \wedge \partial_\sigma \partial_\tau h \right) = 0 ~.
 \end{align}
Thus, the three-point amplitudes involving identical spin-three fields is vanishing, as is expected on general grounds, see e.g. \cite{Benincasa:2007xk}. In order to find non-trivial three-point functions we must consider the even spin case, and rather than proceeding to the spin-four case we now consider the arbitrary higher-spin case.

\section{Generalisation to Higher Spins}

One may generalise the previous discussion to consider deformations corresponding to $(0,1)$-forms taking values in sections of higher-rank symmetric products of the tangent bundle. Namely, we consider the operator
 \begin{align}
	\bar\partial_f = \bar\partial + f^{(n)} ~,
 \end{align}
where 
 \begin{align}
	f^{(n)} = f^{\alpha_1\dots \alpha_n} \partial_{\alpha_1} \cdots \partial_{\alpha_n} 
		\in \Omega^{0,1}(\PTc, \mathrm{Sym}^n(T^{1,0}(\PTc))) ~ .
 \end{align}
The tensor fields $f^{\alpha_1\dots \alpha_n}$ are elements of $\Omega^{0,1}(\PTc, {\cal O}(n))$, and in order to be well-defined on $\PTc$ they have to have the gauge freedom
 \begin{align}
\label{eq:fgauge_n}
	f^{\alpha_1 \dots \alpha_n} \rightarrow  f^{\alpha_1\dots \alpha_n} + Z^{(\alpha_1}\Lambda^{\alpha_2\dots\alpha_n)}~,
 \end{align}
where the gauge parameter is itself a rank $n-1$ tensor, $\Lambda^{(\alpha_2\dots\alpha_{n})}\in \Omega^{0,1}(\PTc, {\cal O}(n-1))$\,. For convenience we will henceforth mostly use a multi-index notation defined as follows: let $I$ be an ordered set $\set{1, \dots, n}$, we then write
 \begin{align}
	f^{\alpha_I} :=f^{\alpha_1\cdots \alpha_n} 
	~~~ {\rm and} ~~~
	\partial_{\alpha_I} := \partial_{\alpha_1} \cdots \partial_{\alpha_n}~,
 \end{align}
so that $f^{(n)} = f^{\alpha_1\dots \alpha_n} \partial_{\alpha_1} \cdots \partial_{\alpha_n} \equiv f^{\alpha_I} \partial_{\alpha_I}$\,. We will denote the degree of the index by $|I|=n$. 

The equations of motion for the $f^{\alpha_I}$ field in the self-dual sector will be
 \begin{align}
	\bar\partial f^{\alpha_I} +  f^{\beta_I} \wedge \partial_{\beta_I} f^{\alpha_I}  = 0 ~.
 \end{align}
As in the spin-three case, we introduce the totally symmetric Lagrange multiplier fields $g^\Omega_{(\alpha_1\dots \alpha_n)}= g_{(\alpha_1\dots \alpha_n)}\otimes\Omega$, where $g_{(\alpha_1\dots \alpha_n)} \in \Omega^{0,1}(\PTc, {\cal O}(-n-4))$\,, and as before $\Omega$ is the weight four holomorphic volume $(3,0)$-form.
The constraints ensuring gauge invariance are now  $Z^{\alpha_1} g_{(\alpha_1\dots \alpha_n)}=0$. We will also denote such fields  using the multi-index notation $g_{(\alpha_1\dots \alpha_n)} = g_{\alpha_I}$\,. 
The twistor action for the self-dual sector can be written as
 \begin{align}
 \label{eq:hs_sd_action}
	S_{\rm s.d.}\left[f^{(n)},g^{(n)}\right] = \int_{\PTc} \Omega \wedge  \left( g_{\alpha_I}\wedge (\bar \partial f^{\alpha_I}+ f^{\beta_J}\wedge \partial_{\beta_J} f^{\alpha_I})\right) ~ .
 \end{align}

\subsection{Linearised Spin-\texorpdfstring{$s$}{\textit{s}} Fields}

At the linearised level the equations of motion are simply $\bar \partial f^{\alpha_I}=0$ and $\bar \partial g_{\alpha_I}=0$\,. Using the Penrose transform, one could find the space-time fields corresponding to the twistors fields $f^{\alpha_I}$ and $g_{\alpha_I}$, and so show that they satisfy the massless spin-$s$ wave equation in exactly the same fashion as in the spin-two and spin-three cases. Instead, as it also allows us to discuss the truncation to the unitary sector, we will briefly consider the on-shell spectrum of the theory before considering the on-shell three-point interactions corresponding to the self-interactions of these spin-$s$ fields.

\paragraph{Minkowski space-time spectrum:} 

We can repeat the analysis of \secref{ssec:Mspec} for this general case. The twistor fields  $f^{\alpha_1\dots \alpha_{n}}(Z)$ have homogeneity $n$, and the fields $g_{\alpha_1\dots \alpha_n}(Z)$ are of homogeneity $-n-4$, while the gauge invariance and constraint are now
 \begin{align}
	f^{\alpha_1 \dots \alpha_n} \rightarrow  f^{\alpha_1\dots \alpha_n}+Z^{(\alpha_1}\Lambda^{\alpha_2\dots\alpha_n)} 
	~~~{\rm and}~~~
	g_{\alpha_1 \dots \alpha_n}Z^\alpha = 0~.
 \end{align}
Following the prescription in \secref{ssec:Mspec} for the tensor $f^{\alpha_1\dots \alpha_{n}}(Z)$ we can form 
 \begin{align}
	\sum_{\ell=0}^{n}(n+1-\ell)(\ell+1)=\tfrac{1}{6}(n+1)(n+2)(n+3)
 \end{align}
invariants, and so after removing the gauge degrees of freedom, for which we repeat the counting above but with $n$ replaced by $n-1$, we have $\tfrac{1}{2}(n+2)(n+1)$ on-shell degrees of freedom. As the highest homogeneity is $2n$, the highest helicity state is $s=1+(2n)/2=n+1$, and so we have  $\tfrac{1}{2}s(s+1)$ degrees of freedom. Using the duality argument discussed above in the spin-three case, we find exactly the same number of on-shell states from $g_{\alpha_1\dots \alpha_n}(Z)$ but with the opposite helicities, and so the total number of on-shell degrees of freedom is 
 \begin{align}
	\nu_s=s(s+1)~.
 \end{align}
For $s=1$ we find the usual number two on-shell vector states while for $s=2$ we find the six on-shell degrees of freedom of Weyl gravity \cite{Riegert:1984hf}. More generally, the formula matches with the number of on-shell states in the conformal higher spin theory described by Fradkin and Tseytlin \cite{Fradkin:1985am}.

\paragraph{A unitary subsector:} 

We can similarly extend the analysis of the diagonalisable sector of on-shell states by simply adding more indices to the spin-three case, and again the on-shell representation of the Poincar\'e generators will fail to be diagonalisable in the full space of conformal spin-$s$ fields. For the case with a non-vanishing cosmological constant and using the infinity twistor, \eqref{eq:infinity_twistor_lambda}, we can define the fields $h_s$ and $\tilde h_s$ corresponding to a subsector for which the generators are diagonalisable by the relations
 \begin{align}
	 \begin{aligned}
	f^{\alpha_1\dots \alpha_n}(Z) &=I^{\beta_1 \alpha_1}\dots I^{\beta_n \alpha_n}\partial_{\beta_1}\dots \partial_{\beta_n} h_{s}(Z)~, \nn \\
	g_{\alpha_1\dots \alpha_n}(Z) &=I_{\alpha_1\beta_1}\dots I_{\alpha_n\beta_n}Z^{\beta_1} \dots Z^{\beta_n}{\tilde h_{s}}(Z) ~,\nn
	 \end{aligned}
 \end{align}
 or using the multi-index notation, where $	I_{\alpha_I\beta_I} := I_{\alpha_1\beta_1} \cdots I_{\alpha_n\beta_n}$,
 \begin{align}
\label{eq:u_ss_n}
	 \begin{aligned}
	f^{\alpha_I}(Z)  = I^{\beta_I \alpha_I} \partial_{\beta_I} h_s(Z) ~, 
	\qquad
	g_{\alpha_I}(Z)  = I_{\alpha_I\beta_I} Z^{\beta_I} \tilde h_s(Z) ~,
	 \end{aligned}
 \end{align}
 as well as $I^{\alpha_I\beta_I} I_{\beta_I\gamma_I} =(- \Lambda)^{\abs{I}} \delta^{\alpha_I}_{\gamma_I}$\,.
In this case, in the linearised approximation about flat twistor space we have
that $h_s\in H^{0,1}(\PT, \calO(2s-2))$ and $\tilde h_s\in H^{0,1}(\PT, \calO(-2s-2))$ so that $h_{s}(Z)$ corresponds to a state of spin $s=(n+1)$ and ${\tilde h_{s}}(Z)$ to a state of $s=-(n+1)$\,. Taking the limit $\Lambda\to 0$ produces the higher-spin analogue of the truncation of conformal gravity to Einstein gravity at the linearised level around Minkowski space-time. That is for every spin-s we have two on-shell degrees of freedom, $h_s$ and $\tilde h_s$, which correspond to space-time helicities of $\pm s$. This is the spectrum of massless higher spins found by Fronsdal \cite{Fronsdal:1978rb} for the spectrum of the massless limit of the Hagen-Singh theory \cite{Singh:1974qz}.

\subsection{Spin-\texorpdfstring{$s$}{s} Three-Point \texorpdfstring{$\overline{\rm MHV}$}{anti-MHV} Amplitudes}

For generic spin-$s$, unlike in the spin-three case, the cubic interactions involving fields corresponding to the same spin do not necessarily vanish. Explicitly focusing on the unitary subsector by substituting the expression \eqref{eq:u_ss_n}, and dropping the subscript on $h$ and $\tilde h$, the three-point interaction reads
 \begin{align}
	S_\text{s.d.}^\text{3pt}\left[f^{(n)},g^{(n)}\right] & = \int_{\PTc} \DD[3]Z \wedge g_{\alpha_I} \wedge f^{\beta_I} \wedge \partial_{\beta_I} f^{\alpha_I} \nn \\
\label{eq: S3pt spin s}
		& = (s-1)! \Lambda^{s-1} \int_{\PTc} \DD[3]Z \wedge \tilde h \wedge \set{h, h}_{s}~,
 \end{align}
where
 \begin{align}
	\set{h,k}_{s} & := (-1)^{s-1} I^{\gamma_I \sigma_I}\partial_{\gamma_I} h \wedge \partial_{\sigma_I} k 
	~~~ \text{for } \abs{I} = s-1~.
 \end{align}
Clearly, $\set{h,k}_{s} = (-1)^{s} \set{k,h}_{s}$ and thus $\set{h,h}_{s} = 0$ for any odd integer $s$, which implies that the $\overline{\rm MHV}$ three-point amplitude is vanishing for any odd spin.

The non-vanishing even-spin three-point $\bar{\rm MHV}$ amplitudes can be easily extracted from \eqref{eq: S3pt spin s} with an appropriate choice of wavefunctions. We choose plane wave momentum eigenfunctions following \cite{Witten:2004cp}, see also \cite{Adamo:2013cra}, where we allow the space-time momenta to be complex. These can be written using the definition of the delta-function on the complex plane as a $(0,1)$-form
 \begin{align}
	\bar \delta(a z-b)=\frac{1}{2\pi i} \dd\bar z \frac{\partial}{\partial \bar z}\left(\frac{1}{a z-b}\right)~.
 \end{align}
In particular, this satisfies for a holomorphic function $f(z)$
 \begin{align}
	\int \dd z\wedge \bar \delta(a z- b ) f(z)=\frac{1}{a}f(\tfrac{b}{a})~.
 \end{align}
By considering spinors $\lambda_A=(1,z)$ and $\lambda'_A=(b,a)$ we can rewrite this as:
 \begin{align}
	\bar \delta(\langle \lambda\lambda'\rangle)= \frac{1}{2\pi i} \dd\bar \lambda^{\dot A}\frac{\partial}{\partial \bar \lambda^{\dot A}}\frac{1}{\langle \lambda\lambda'\rangle} ~.
 \end{align}
This object has homogeneity $-1$ in both $\lambda$ and $\lambda'$, while to define wavefunctions for higher-spin particles it is necessary to consider different scalings. To this end we introduce an arbitrary reference spinor $\xi$ and define
 \begin{align}
 \label{eq:weight_m_delta}
	\bar \delta_m(\langle \lambda\lambda'\rangle)= \left(\frac{\langle  \xi \lambda\rangle}{\langle \xi \lambda'\rangle}\right)^m\bar \delta(\langle \lambda\lambda'\rangle)~,
 \end{align}
which has homogeneity $m-1$ in $\lambda$ and $-m-1$ in $\lambda'$. Alternatively by introducing a complex parameter $u$ and another $(0,1)$-form, $\bar\delta(u- \tfrac{\langle  \xi \lambda'\rangle}{\langle \xi \lambda \rangle})$, we can, after a change of integration variables, write \eqref{eq:weight_m_delta} as 
 \begin{align}
	\bar \delta(\lambda, \lambda')=\int \frac{\dd u}{u^m}\wedge \bar \delta^{2}( u \lambda-\lambda'  )~.
 \end{align}
We use these weighted delta-functions to define our plane wave momentum eigenfunctions describing particles with on-shell momenta given by the spinors $(p_i, \tilde p_i)$ (in the following the subscripts do not denote the spin, but label the particles)
 \begin{align}
	 h_i(Z) & = \int_\Complex \frac{\dd u_i}{u_i}\frac{1}{u_i^{2s-2}}\wedge \bar\delta^{2}( u_i\lambda - p_i) e^{u_i \chirm{\mu \tilde p_i}}~, \nn\\
 \tilde h_i(Z) & = \frac{1}{(s-1)!}\int_\Complex \frac{\dd u_i}{u_i} \frac{1}{u_i^{-2s-2}} \wedge \bar\delta^{2}( u_i\lambda - p_i) e^{u_i \chirm{\mu \tilde p_i}}~. \nn
 \end{align}
The (0,1)-forms $h$ and $\tilde h$ have homogeneity $2s-2$ and $-2s-2$ respectively under $Z=(\lambda, \mu) \rightarrow t Z$, while under the rescaling of space-time momentum helicity spinors, $p_i\rightarrow t p_i$, $\tilde p_i \rightarrow t^{-1} \tilde p_i$, they scale as $t^{2s}$ and $t^{-2s}$ as is expected. We can now compute the three-point $\bar{\rm MHV}$ amplitude.

As previously mentioned the odd spin $s$ amplitude obviously vanishes. The three-point $\bar{\rm MHV}$ amplitude for even spin $s$ is given by 
 \begin{align}
	& \calM_{3}(-s, +s, +s) = \Lambda^{s-1} (s-1)! \int_{\PTc} \DD[3]Z \wedge \tilde h_1 \wedge \set{h_2, h_3}_{s}\nn\\
		& \qquad = (-\Lambda)^{s-1} \int_{\PTc} \DD[3]Z \wedge \tilde h_1 \wedge \left(\vphantom{\partiald{}{\lambda_A}} \right.\nn\\
		& \qquad\qquad \,\phantom{+}\, \Lambda^{s-1} \epsilon_{A_1 B_1}\cdots \epsilon_{A_{s-1}B_{s-1}} \partiald{}{\lambda_{A_1} }\cdots \partiald{}{\lambda_{A_{s-1}}} h_2 \wedge \partiald{}{\lambda_{B_1}} \cdots\partiald{}{\lambda_{B_{s-1}}} h_3 \nn \\
		& \qquad\qquad + (s-1) \Lambda^{s-2} \epsilon^{A_1'B_1'}\epsilon_{A_2B_2} \cdots \partiald{}{\mu^{A_1'}} \partiald{}{\lambda_{A_2}} \cdots h_2 \wedge \partiald{}{\mu^{B_1'}}\partiald{}{\lambda_{B_2}} \cdots  h_3 \nn \\
		& \qquad\qquad + \dots \nn \\
		&\left. \qquad\qquad +\, \epsilon^{A_1'A_2'} \cdots \epsilon^{A_{s-1}'B_{s-1}'} \partiald{}{ \mu^{A_1'} } \cdots \partiald{}{ \mu^{A_{s-1}'} }  h_2 \wedge \partiald{}{ \mu^{B_1'} }\cdots \partiald{}{ \mu^{B_{s-1}'} } h_3\right)\,. 
 \end{align}
We can now use the fact that 
 \begin{align}
	\partiald{h_2}{\mu^{A'}}\wedge \partiald{h_3}{\mu_{A'}} 
		= \chirm{32} \int \frac{\dd u_2\wedge \dd u_3 }{(u_2 u_3)^{2s -2} }\wedge 
			\bar \delta (u_2 \lambda-p_2) \wedge  \bar \delta (u_3 \lambda-p_3) e^{\sum_{i=2}^3 u_i[\mu, \tilde p_i]}\nn
 \end{align}
and 
 \begin{align}
	\partiald{h_2}{\lambda_{A}} \wedge \partiald{h_3}{\lambda^{A}} 
		= \chirp{\partial_2 \partial_3 }\int \frac{\dd u_2\wedge \dd u_3 }{(u_2 u_3)^{2s -2} }\wedge
			\bar \delta (u_2 \lambda-p_2) \wedge  \bar \delta (u_3 \lambda-p_3) e^{\sum_{i=2}^3 u_i[\mu, \tilde p_i]}\nn
	~,
 \end{align}
where the bracket $[23]$ involves space-time momenta and is defined in general as $[ij]=\epsilon^{A'B'}{\tilde p}_{i A'}{\tilde p}_{j B'}$\,, and similarly
 \begin{align}
 \chirp{\partial_i \partial_j }= \epsilon_{AB}\partiald{}{p_{i A}}\partiald{}{p_{j B}}\, .
 \end{align}
We can thus write the three-point amplitude as
 \begin{align}
	\calM_{3}(-s, +s, +s) = \Lambda^{s-1} \left(\Lambda \chirp{\partial_2 \partial_3} + \chirm{23}\right)^{s-1}{ \cal I} ~,
 \end{align}
where the integral ${\cal I}$ is given by 
 \begin{align}
	{\cal I} &= \frac{1}{\chirm{23}}\int \DD[3] Z \wedge \dd u_1\wedge \dd u_2 \wedge \dd u_3\wedge \prod_{i=1}^3 \bar \delta^{2}(p_i-u_i \lambda) 
			\frac{u_1^{2s+1}}{(u_2u_3)^{2s}} e^{\sum_{i=1}^3u_i[\mu,\tilde p_i]}\nn
	~,
 \end{align}		
The integration over the background projective twistor space gives a delta-function on whose support $u_2 = \chirm{31}/\chirm{23}$ and $u_3 = \chirm{12}/\chirm{23}$\,. The remaining integrals are trivial, and a factor $\chirm{23}$ comes from the Jacobian of the transformation to the usual expression of the four-dimensional momentum delta-function. Hence we find
\begin{align} 
		{\cal I}&=\chirm{23}\left(\tfrac{\chirm{2 3}^{2}}{\chirm{12} \chirm{31}}\right)^{s-1} \delta^{4}(P) ~,
 \end{align}
where $P=\sum_{i=1}^3p_i$ is the total momentum. Finally, using the fact that 
 \begin{align}
	\chirp{\partial_2 \partial_3 } \delta^{4}(P)&= -\chirm{23} \Box_P \delta^{4}(P)
 \end{align}
with $\Box_P=-\tfrac{1}{2} \epsilon_{AB}\epsilon_{A'B'} \partiald{}{P_{AA'}} \partiald{}{P_{BB'}}$\,, 
we obtain
 \begin{align}
\label{eq: 3pt amp s}
	\calM_{3,-1}(-s, +s, +s) & = \begin{cases} \Lambda^{s-1} \left(\frac{\chirm{2 3}^{3}}{\chirm{12} \chirm{31}}\right)^s \left(1 - \Lambda \dAl_P\right)^{s-1} \delta^{4}(P) 
 & \text{ for $s$ even} \\
	0 &\text{ for $s$ odd~.} \end{cases}
 \end{align}
These amplitudes vanish in the flat space limit $\Lambda\to 0$, however we can define a rescaled amplitude $\Lambda^{-(s-1)} \calM_{3}$ to obtain a non-vanishing result at least for even spins. This expression coincides with the result obtained from general space-time symmetry arguments for massless higher-spin three-point amplitudes \cite{Benincasa:2007xk}.

\section{Toward Consistent Higher Spin Interactions}
\label{sec:cons_ints}

There are well-established reasons for believing that there is no consistent interacting theory for a single $s>2$ field, and in the twistor theory we can see such difficulties emerge when we attempt to give a geometric interpretation to our equations of motion.
As we have mentioned, beyond the spin-two case our equation of motion does not imply that the operator $\bar \partial_f$ is integrable. This failure can be see quite easily in the spin-three case where
 \begin{align}
	\bar \partial_f^{\,2} = \left( \bar \partial f^{\alpha\beta}+f^{\gamma\delta}\wedge \partial_{\gamma}\partial_{\delta} f^{\alpha\beta}\right) \partial_\alpha\partial_\beta + 2 f^{\gamma \delta}\wedge \partial_\gamma f^{ \alpha\beta}\partial_{\delta} \partial_\alpha \partial_\beta~,
 \end{align}
with the last term remaining even after imposing the spin-three equations of motion previously used. However, the structure of this term involving three derivatives appears in the spin-4 equation, and so one can include this additional term as a source term for spin-4 fields. Of course the spin-4 fields will generate additional terms which will not be cancelled but which in turn source spin-5 fields and so on. 

It is not difficult to also include spin-one fields in our description, though they have to be treated somewhat differently as they do not take values in some product of the tangent bundle but in the Lie algebra $\alg{g}$ corresponding to a given Lie (gauge) group $G$, i.e. $f^{\alpha_\emptyset}\equiv A^+ \in \Omega^{0,1}(\PTc, \calO(0) \otimes \alg{g})$ and $g_{\alpha_\emptyset}\equiv A^- \in \Omega^{0,1}(\PTc, \calO(-4) \otimes \alg{g})$\,. To avoid unnecessary clutter like taking traces when necessary, we will just consider the abelian case $G = U(1)$, however a generalisation is straightforward. Thus, whenever we write $\alg{g}$ we just consider $\alg{u}(1)$ for simplicity. We can now define the operator
 \begin{align}
\label{eq: dbarf HS general}
	\bar\partial_f= \bar\partial + \sum_{\abs{J}=0}^{\infty} f^{\beta_J}\partial_{\beta_J}~.
 \end{align}
We can expand the condition $\bar \partial_f^{\,2}=0$ in powers of derivatives $\partial_{\alpha}$ and impose the vanishing component by component. Focusing on spin two we have
 \begin{align}
	\bar \partial f^\alpha+\sum_{\abs{J}=0}^\infty f^{\beta_J} \wedge \partial_{\beta_J} f^\alpha=0 ~,
 \end{align}
where we see the coupling of the spin-two field to all higher-spin fields. In this equation it is of course consistent to set all the $f^{\alpha_I}$ for $|I|>1$ as well as $\abs{I} = 0$ equal to zero and so recover the spin-two equation of motion for pure conformal gravity. For the spin-three equation of motion we now have
 \begin{align}
	\bar \partial f^{\alpha_1\alpha_2}+\sum_{\abs{J}=0}^\infty f^{\beta_J} \wedge \partial_{\beta_J} f^{\alpha_1\alpha_2}
+\sum_{\abs{J}=0}^\infty (\abs{J}+1)f^{(\alpha_1 \beta_J} \wedge \partial_{\beta_J} f^{\alpha_2)}=0 ~.
 \end{align}
Because $f^{(\alpha_1}\wedge f^{\alpha_2)}=0$\,, the spin-three fields are not sourced by purely spin-two fields, and so truncating to just spin-two is consistent as is expected. For generic spin, and using the multi-index notation, we have the equation  
 \begin{align}
 \label{eq:MC_arb_spin}
	N^{\alpha_I}\equiv\bar \partial f^{\alpha_I}+\sum_{\abs{J}=0}^{|I|}\sum_{|K|=0}^\infty C_{{|K|}{|J|}} f^{(\alpha_J\beta_K} \wedge \partial_{\beta_K} f^{\alpha_{I-J})}=0~,
 \end{align}
where the multi-index $I-J$ corresponds to the complement of $J$ in $I$, and the coefficients $C_{{\abs{K}}{\abs{J}}} = {\binom{\abs{K}+\abs{J}}{\abs{J}}}$\,. Here we can see the source terms due to lower-spin fields in higher-spin equations of motion, and for example as the spin-4 equation involves non-vanishing source terms from the spin-three field, we cannot truncate to just the spin-three sector. Hence we see the need for an infinite number of fields, one of each spin, interacting non-trivially with one another.

As before, we can introduce Lagrange multiplier fields to impose these conditions to write an action for the self-dual sector
 \begin{align}
 \label{eq:sd_act_ao}
	S_{\rm s.d.}\left[f^{(\bullet)},g^{(\bullet)}\right] = \int_{\PTc} \Omega \wedge  \sum_{|I|=0}^\infty \big( g_{\alpha_I}\wedge N^{\alpha_I}\big) ~.
 \end{align}
The equation of motion for the fields $g^{(n)}$ following from this action can be directly derived. At the linearised level the analysis is as in the previous sections as the individual spins decouple, moreover one can focus on the individual spins to calculate the self-interaction three-point functions, and the results from previous sections will still hold. Nonetheless, this action is significantly more involved, and without a proper geometric understanding of the deformation much of the structure remains unclear. 

One natural approach is to attempt to interpret the deformation as defining a new complex structure. For example we can consider the space spanned by the deformed vectors $e_{ \alpha}= \partial_{ \alpha} + \sum_{\abs{I}=0}^\infty \bar f_{ \alpha}^{\bar \alpha_I}\bar \partial_{\bar \alpha_I}$ and ask if it is closed under commutation. That is given $V=V^{ \alpha}e_{ \alpha}$ and $W=W^{ \alpha}e_{ \alpha}$ we calculate
 \begin{align}
	[V,W] &= \left( V^{ \alpha }( \partial_{ \alpha}^{\bar f} W^{ \beta})-W^{ \alpha }( \partial_{ \alpha}^{\bar f} V^{ \beta}) \right) e_{ \beta} \nn\\
		& \qquad + \left(\sum_{|I|=0}^\infty \sum_{|J|=0}^{|I|} C_{{|I|}{|J|}}\left( V^\alpha \bar \partial_{\bar \gamma_J}W^\beta -W^\alpha \bar \partial_{\bar \gamma_J}V^\beta\right) {\bar f}_\alpha^{\bar \gamma_J\bar \gamma_{I-J}} \bar \partial_{\bar \gamma_{I-J}}\right) e_\beta~.
 \end{align}
The presence of the second line would seem to require at the very least a significant generalisation of the usual notions. In particular the appearance of infinite numbers of derivatives suggests a non-local formulation. This can also be seen if we consider a $C^\infty$-function on projective twistor space $\phi(Z)$ and attempt to define the notion of holomorphicity with respect to a deformed complex structure by writing
 \begin{align}
	\bar \partial_f \phi(Z)=\bar \partial \phi(Z) +\sum_{|I|=0}^\infty f^{\alpha_I}\partial_{\alpha_I}\phi(Z)=0~.
 \end{align}
This condition we can be written in a more suggestive notation as 
 \begin{align}
 \label{eq:hol_cond}
	\bar \partial \phi(Z) +f\cdot \phi(Z)=0
 \end{align}
where 
 \begin{align}
	f\cdot \phi(Z)=\int \DD[3] Z'\, f(Z,Z')\phi(Z')
 \end{align}
with $
	f(Z,Z')=\sum_{\abs{I}=0}^\infty  f^{\alpha_I}(Z) \,\partial_{\alpha_I}^{(Z)}\bar \delta^{3}(Z,Z')$
\text{\footnotemark}%
\footnotetext{
	Here $\bar \delta^{3}(Z,Z')$ is the projective delta-function defined on the background projective twistor space such that 
	 \begin{align}
		\phi(Z)=\int \DD[3] Z\wedge \bar \delta^{3}(Z,Z') \phi(Z')~.
	 \end{align}
	}%
	\,.
Such bi-local expressions are common in higher spin theories and have been interpreted in terms of infinite jet bundles \cite{Leigh:2014tza, Leigh:2014qca}%
\footnote{
	It is interesting to note that he unfolded formalism of higher spin theory also has a natural interpretation in terms of jet spaces, see \cite{Taronna:2016ats} for a recent discussion and references.
	}%
.

\subsection{A Geometric Interpretation of Higher Spins}

The language of jet bundles,  which we briefly review below, will allow us to give a more geometric interpretation of the higher-spin equations of motion as the integrability condition for a holomorphic structure.  Given a complex manifold, here we are obviously considering $\PTc$, we can define the corresponding Dolbeault operator $\bar{\partial}$ mapping $(p,q)$-forms to $(p,q+1)$-forms. This can be naturally generalised to $(p,q)$ forms taking values in sections of some complex vector bundle $B\rightarrow \PTc$, that is elements of $\Omega^{p,q}(\PTc; B)$.
We will be mostly considering bundles whose sections are symmetric covariant tensors of $\mathrm{Sym}^n(T\PTc)$ or contravariant tensors of $\mathrm{Sym}^n(T^*\PTc)$ or $\alg{g}$-valued for $n=0$\,. We wish to define a holomorphic structure on $B$, this is a sequence of operators 
 \begin{align}
	\bar{\partial}_{\cal B}:\Omega^{p,q}(\PTc, B)\rightarrow \Omega^{p,q+1}(\PTc, B)
 \end{align}
such that 
 \begin{align}
	{\rm i)}~ &~~\bar{\partial}_{\cal B} \circ \bar{\partial}_{\cal B}=0\nn\\
	{\rm ii)}~ &~~ \bar{\partial}_{\cal B}(\omega \wedge g)=\bar{\partial}(\omega) \wedge g+(-1)^{m+n}\omega\wedge \bar{\partial}_{\cal B}(g)
 \end{align}
where $\omega \in \Omega^{m,n}(\PTc)$ and $g \in \Omega^{p,q}(\PTc, B)$ for any $(m, n)$ and $(p,q)$. 

The holomorphic structure $\bar{\partial}_{\cal{B}}$ on $B$ induces a holomorphic structure, also denoted $\bar{\partial}_{\cal{B}}$, on ${\rm End}(B)$, 
 \begin{align}
	\bar{\partial}_{\cal{B}}(f)=\bar{\partial}_{\cal{B}} \circ f-(-1)^{p+q} f \circ \bar{\partial}_{\cal{B}}~, ~~~f \in \Omega^{p,q}(\PTc, \rm{End}(B))~.
 \end{align}
Let $\bar{\partial}_{{\cal B}'}$ be another holomorphic structure on $B$, then there exists a section $f\in \Omega^{(0,1)}(\PTc, {\rm End}(B))$ such that
 \begin{align}
\label{eq:def_hol_struc}
	\bar{\partial}_{{\cal B}'}(g)=\bar{\partial}_{{\cal B}}(g)+f\circ g 
 \end{align}
where $g \in \Omega^{p,q}(\PTc, B)$ and $f$ satisfies the Mauer-Cartan equation 
 \begin{align}
\label{eq:MC}
	\bar{\partial}_{{\cal B}}f+f\circ f=0~.
 \end{align}
Conversely if $f$ is a section satisfying \eqref{eq:MC} then defining $\bar{\partial}_{{\cal B}'}$ by \eqref{eq:def_hol_struc} gives another holomorphic structure on $B\rightarrow \PTc$. 

Making contact with our previous considerations we see that the deformation in \eqref{eq: dbarf HS general}, $f$, is to take values in $\Omega^{0,1}(\PTc, {\rm Sym}^{(\bullet)}(T^{1,0}\PTc))$, i.e. $f(Z) =\sum f^{\alpha_I}_{\bar\alpha}(Z) \dd{\bar Z}^{\bar\alpha} \partial_{\alpha_I}$. For $\abs{I}>1$, this is not a derivation and therefore does not give rise to a holomorphic structure in the usual sense. In order to interpret the higher powers of derivatives as linear operators on some vector space we must think of the deformed operators as acting on an infinite dimensional vector formed from the field $\phi$ and all of its derivatives 
 \begin{align}
 	(\phi, \partial_{\alpha} \phi, \partial_{\alpha_1}\partial_{\alpha_2}\phi, \dots) ~.
 \end{align}
Such an object is an infinite jet called the infinite prolongation $j^\infty \phi$ of $\phi$ and and we can now interpret $f$ as taking values in the endomorphisms of the jet bundle. The higher powers of derivatives act as generators of the space of endomorphisms.

\paragraph{Jet bundles:} 

To be slightly more precise, see \cite{saunders1989geometry} for a textbook treatment, we wish to consider fields which are sections of some bundle $B$ i.e. the fields will be sections of $\mathrm{Sym}^{(n)}(T^{1,0}\PTc)$ or $\mathrm{Sym}^{(n)}(T^{*}{}^{1,0}\PTc)$ or $\alg{g}$. To describe this we choose an appropriate adapted local coordinate system, $\psi$, on the total space of the bundle $B$ where for a given subspace $W\subset B$ the coordinates can be split into those parametrising the base space and those distinguishing points on the fibre: $\psi=(Z^\alpha, \psi_{\beta_J})$.
Given two such sections, say $g$  and $\tilde g$, we say they have the same $k$-jet at $Z\in \PTc$ if in any particular coordinate system their first $k$ derivatives coincide, i.e. 
 \begin{align}
	\partial_{\alpha_I} g_{\beta_J}(Z)= \partial_{\alpha_I} \tilde g_{\beta_J}(Z)\, , 
	~~~ 0\leq \abs{I}\leq k~,
 \end{align}
where $g_{\beta_J}=\psi_{\beta_J}\circ g$. This definition is in fact independent of the particular coordinate system. The $k$-th jet of $g$ at $Z$, denoted $j^k_Zg$, is the equivalence class of all sections with the same $k$-jet. The $k$-th jet manifold, which we denote $J^k(B)$, is the totality of all such jets, 
 \begin{align}
	J^k(B) = \left\{j^k_Z g:\forall  Z\in \PTc, g \in  \Omega^{0,1}(\PTc,B) \right\} \,.
 \end{align}
The jet-manifold combined with the so-called source projection $\pi_k: j^k_Z g\rightarrow Z$, can be viewed as a bundle over the base space $\PTc$. 
The coordinate system $\psi$ on $B$ induces a coordinate system $\psi^k$ on $J^k(B)$: given $W\subset B$ we define
$W^k=\{j_Z^k g:g(Z)\in W\}$ and 
 \begin{align}
	\psi^k=(Z^\alpha, \psi_{\beta_J}, \psi_{\beta_J; \alpha_1}, \dots, \psi_{\beta_J; \alpha_1\dots \alpha_k})
 \end{align}
are called derivative coordinates where for $j^k_Zg\in W^k$ we define $Z^\alpha(j^k_Zg)=Z^\alpha$, $\psi_{\beta_J}(j^k_Zg)=g_{\beta_J}(Z)$, and 
 \begin{align}
	\psi_{\beta_J;\alpha_I}(j^k_Zg)=\partial_{\alpha_I}g_{\beta_J}(Z)~.
 \end{align}
Correspondingly, given a open subset $U\in \PTc$ and local section $g\in \Gamma_U(B)$ we define the $k$-th prolongation of $g$ as the section $j^kg \in \Gamma_U(J^k(B))$ defined by $j^k g(Z)=j_Z^k g$ for $Z\in U$ with coordinate representation
 \begin{align}
	(g_{\beta_J}, \partial_{\alpha_1}g_{\beta_J}, \dots, \partial_{\alpha_1}\dots \partial_{\alpha_k}g_{\beta_J})~.
 \end{align}
It is worthwhile to note that, while these prolongations will be the focus of our interest, they are very non-generic sections of the jet bundle since their adaptive coordinates are strongly related to each other, which generically does not need to be the case. The infinite jet bundle corresponds to the limiting case $k\to \infty$. For $Z\in \PTc$ the $\infty$-th jet of $g$, which we denote $j^\infty_Z g$, is the equivalence class of sections whose derivatives coincide with those of $g$ at all orders and the space $J^\infty (B)$ is an infinite dimensional manifold which can be shown to have the structure of a bundle over $\PTc$.

We can now view $f$ as defining a bundle endomorphism on $J^\infty(B)$. In the example where we are only considering $C^\infty$ functions, as in \eqref{eq:hol_cond}, we can still construct the corresponding infinite jet bundle, denoted simply $J^\infty(\PTc)$,  and we can replace the action of the deformation $f$ on a function $g$ at a point $Z$ by its action on the corresponding prolongation $j^\infty( g) \in \Gamma(J^\infty(\PTc))$
 \begin{align}
 \label{eq:jet_lop_1}
 	(f\cdot g)(Z) \to 	f \cdot j^\infty_Z( g) 
 		& \equiv \sum_{\abs{I}=0}^\infty j^\infty_Z (f^{\alpha_I}) \wedge \T\indices{_{;\alpha_I}}[ j^\infty_Z (g)] 
		= \sum_{\abs{I}=0}^\infty  j^\infty_Z (f^{\alpha_I} \wedge \partial_{\alpha_I}g)
 \end{align}
where $\T\indices{_{;\alpha_I}} \in\mathrm{End}(J^\infty(\PTc))$ such that $\T\indices{_{;\alpha_I}} [j^\infty_Z( g)]= j^\infty_Z (\partial_{\alpha_I} g)$\,, and where in the wedge product it is understood that we use the product rule for jets $j^\infty_Z(g_1) \cdot j^\infty_Z(g_2)=j^\infty_Z(g_1  g_2)$. In an adapted coordinate system this product can be given explicitly by using the formula for the generalised higher order Leibniz rule. We note that the action of the generators $\T\indices{_{;\alpha_I}}$, despite appearances, is linear and
 \begin{align}
	\T\indices{_{;\alpha_I}}[ \omega(Z) j^\infty_Z (g)] 
		= \omega(Z) \T\indices{_{;\alpha_I}}[  j^\infty_Z (g)] 
		= \omega(Z) j^\infty_Z (\partial_{\alpha_I} g)
 \end{align}
for $\omega(Z)$ being an arbitrary function as the derivatives acting on $g$ only appear due to the particular structure of the prolongation.

If we consider $j^\infty(g)\in \Omega^{(p,q)}(\PTc, J^\infty)$, i.e. $(p,q)$-forms taking values in $J^\infty(\PTc)$, we can now define a corresponding holomorphic structure
\begin{align}
	\bar \partial_f(j^\infty(g)) & \equiv \bar{ \partial} j^\infty(g) + f\cdot j^\infty(g)\nn\\
		&=\bar{ \partial} j^\infty(g) + \sum_{\abs{I}=0}^\infty j^\infty(f^{\alpha_I})\wedge \T\indices{_{;\alpha_I}}[  j^\infty(g)]\nn\\
		&= j^\infty\big(\bar{ \partial} g + \sum_{\abs{I}=0}^\infty f^{\alpha_I}\wedge \partial_{\alpha_I} g\big)~.
 \end{align}
As $\T\indices{_{;\alpha_I}}[\omega\wedge  j^\infty(g)]=\omega \wedge \T\indices{_{;\alpha_I}}[  j^\infty(g)]$ for $\omega \in \Omega^{m,n}(\PTc)$ we have that $f\cdot (\omega\wedge j^\infty(g))=(-)^{(m+n)} \omega \wedge f\cdot  j^\infty(g)$\,. Moreover we see that 
 \begin{align}
	\bar \partial_f \circ \bar \partial_f=0 
	\quad \Longleftrightarrow \quad 
	j^\infty\big(\sum_{\abs{I}=0}^\infty \bar{\partial} f^{\alpha_I} \wedge \partial_{\alpha_I}g 
				+ \sum_{\abs{I}=0}^\infty f^{\beta_I}\wedge \partial_{\beta_I} \big(\sum_{\abs{J}=0}^\infty f^{\gamma_J}\wedge \partial_{\gamma_J}g \big)\big) = 0 
 \end{align}
for an arbitrary $(p,q)$-form $g$. Hence we see that imposing the conditions \eqref{eq:MC_arb_spin} at all points on $\PTc$ is equivalent to the integrability condition for the holomorphic structure on the infinite jet bundle.

\paragraph{Anti-self-dual fields:}

In addition to the integrability conditions we have equations of motion for the $(0,1)$-form Lagrange multiplier fields $g^\Omega_{\alpha_I} = g_{\alpha_I}\wedge\Omega$. The variation of \eqref{eq:sd_act_ao} with respect to the $f^{\alpha_I}$ fields gives the equations of motion
\allowdisplaybreaks[0]
 \begin{align}
	\bar{\partial}g^\Omega_{\alpha_I}
		- \sum_{\subalign{\abs{J}&=0\\ \abs{K}&=0}}^\infty (-1)^{\abs{K}} & C_{\abs{J}\abs{K}} 
			\partial_{\beta_K} \left(g^\Omega_{\alpha_I\gamma_J}\wedge f^{\gamma_J\beta_K}\right) \nn\\[-4mm]
		& + \sum_{\abs{J}=0}^\infty \sum_{K=0}^{\abs{I}} C_{ \abs{I-K} \abs{K} }
			g^\Omega_{\gamma_J (\alpha_K}\wedge\partial_{\alpha_{I-K})}f^{\gamma_J} = 0~.
  \label{eq:g_eom_1}
 \end{align}
\allowdisplaybreaks

It is worthwhile to note that it is not possible to truncate the theory to just the spin-two case as the lower-spin fields can source the higher-spin ones via the last term. 

We can also interpret the equations \eqref{eq:g_eom_1} in the language of infinite jet bundles, $J^\infty(B)$. As the equations mix fields with different spin we will take 
 \begin{align}
	B= \left( \alg{g} \oplus \bigoplus_{r=1}^\infty{\rm Sym}^{(r)}(T^*{}^{1,0}\PTc) \right) \otimes \Omega^{(3,0)}\otimes {\cal O}(4)~ .
 \end{align}
We define two linear operators which generalise those in \eqref{eq:jet_lop_1}
 \begin{align}
 \label{eq:jet_lop_2}
 \begin{aligned}
	\T_{\gamma_J;\beta_L} g_{\alpha_I}^\Omega & =\partial_{\beta_L}g^\Omega_{\alpha_I\gamma_J}\\
	\T^{\beta_K}_{\gamma_L} g_{\alpha_I}^\Omega & = 
			\begin{cases} 
				0 \quad \text{for } \abs{K}>\abs{I} \text{ or } \abs{K}\geq \abs{I} \text{ and }\abs{L}=0 \\
				\binom{\abs{I}}{\abs{K}} \delta^{\beta_K}_{(\alpha_K} g^\Omega_{\alpha_{I-K}) \gamma_{L}}
					\quad \text{else}\,.
			\end{cases}
 \end{aligned}
 \end{align}

We have compressed the notation by denoting the prolongation $j^\infty g_{\gamma_K}$ by the section $g_{\gamma_K}$. We can now write the equation of motion as 
 \begin{align}
 \label{eq:g_eom_2}
	\bar{\partial}_f g_{\alpha_I}^{\Omega}=\bar{\partial}g^\Omega_{\alpha_I} + F\wedge g^\Omega_{\alpha_I}=0
 \end{align}
where we have introduced the $(0,1)$-forms 
 \begin{align}
 \label{eq:F_jet_connection}
	F & = \sum_{\subalign{\abs{J}&=0\\ \abs{K}&=0}}^\infty \left( \sum_{\abs{L}=0}^{\abs{K}} (-1)^{\abs{K}+1} C_{\abs{J}\abs{K}}C_{\abs{K-L}\abs{L}}\partial_{\beta_{K-L}}
	 f^{\gamma_J\beta_K} \T_{\gamma_J;\beta_L}{}\!
		+ \partial_{\beta_K}f^{\gamma_J} \T_{\gamma_J}^{\beta_K}{}\! \right) ~\nn\\
		&\equiv  \sum_{\subalign{\abs{J}&=0\\ \abs{K}&=0}}^\infty \left( F^{\gamma_J;\beta_K}\T_{\gamma_J;\beta_K}{}\! + F_{\beta_K}^{\gamma_J}\T^{\beta_K}_{\gamma_J}{}\! \right)~.
 \end{align}

\paragraph{Unitary subsector equations of motion:} 

These equations simplify when we focus on the case where the background is conformally flat and the fields are restricted to the diagonalisable sector \eqref{eq:u_ss_n}. For this, $F$ truncates to
 \begin{align}
 \label{eq:H_jet_connection}
	H = \sum_{\subalign{\abs{J}&=0\\ \abs{K}&=0}}^\infty 
			I^{\lambda_J \gamma_J } \left( (-1)^{\abs{K}+1} C_{\abs{J}\abs{K}} 
			I^{\kappa_K \beta_K} \partial_{\lambda_J} \partial_{\kappa_K} h_{\abs{J + K}+1} \T_{\gamma_J;\beta_K}{}\!
			+ \partial_{\beta_K} \partial_{\lambda_J} h_{\abs{J}+1} \T_{\gamma_J}^{\beta_K}{}\! \right)\,. \nn
	\\[-4mm]
 \end{align}

If we further only allow self-interactions of spin $s$ fields, this selects $\abs{J} = 0$ and $K = I$ in the first term, and $J = I$ and $K = I$ in the second term. Hence, we obtain
 \begin{align}
 	H_s = (-1)^{s} I^{\alpha_I \beta_I} \partial_{\alpha_I} h_s \T_{;\beta_I}{}\! 
 				+ I^{\alpha_I \beta_I} \partial_{\alpha_I} \partial_{\gamma_I} h_s \T_{\beta_I}^{\gamma_I}{}\!
 \end{align}
The equations of motion for purely unitary spin-$s$ fields then read
 \begin{align}
	0 & = \left( \bar\partial + H_s \right) \wedge \left( I_{\alpha_I \beta_I} Z^{\beta_I} \tilde h^\Omega_s \right) 
 \end{align}
which equivalently can be written as
 \begin{align}
	0 = Z_{\alpha_I} \bar\partial \tilde h^\Omega_s 
			& - \sum_{\abs{J}=0}^{s-1} 
				\abs{J}! (C_{\abs{I-J} \abs{J}})^2
				Z_{\alpha_{I-J}} \set{ \partial_{\alpha_J} h_s, \tilde h^\Omega_s }_{s - \abs{J}}
			 \nn
 \end{align}
where $Z_{\alpha_J} = Z^{\beta_J} I_{\beta_J \alpha_J}$\,.

\subsection{Anti-Self-Dual Interaction Terms}
\label{sec:asd_int_act}

To go beyond the self-dual sector to the full theory we must include interactions of the anti-self-dual fields. There are a number of possible interactions, however we will restrict ourselves to the simplest case by formulating the twistor analogue of interaction term in \eqref{eq:spin3action}. In this we will closely follow the discussion in \cite{Mason:2005zm}.  

Given a curved twistor space, $\Tc$, with fibre coordinates $\sigma_A$ over a manifold ${\cal M}$ with space-time coordinates $x^{AA'}$, we can choose an adapted vector bundle coordinate system for the cotangent bundle. This defines a set of dual sections in $\Omega^{(1,0)}(\Tc)$ which we label $e^\alpha=(e_A, e^{A'})$. The one-forms $e^A$ are of homogeneity degree one, and when restricted to constant $x^{AA'}$, that is to the fibres of $\Tc\rightarrow {\cal M}$, they are given by $e^A= \dd \sigma^A$. The one-form $e_0=\sigma^A e_A$ is well defined on $\PTc$ with values in ${\cal O}(2)$. The $(1,0)$-forms $e^{A'}$, also of homogeneity one in $\sigma_A$, can be defined at each point to be orthogonal to the fibres of $\Tc\rightarrow {\cal M}$. We can additionally  choose the holomorphic volume form to be
 \begin{align}
	\Omega=\frac{1}{2}\epsilon_{A'B'}e^{A'}\wedge e^{B'}\wedge \sigma_A \dd\sigma^A~.
 \end{align}
In flat twistor space these forms can be given explicitly as
 \begin{align}
	e_A=\dd\sigma_A
	~~~{\rm and} ~~~ 
	e^{A'}=i \sigma_A \dd x^{AA'}
 \end{align}
as well as  
 \begin{align}
	\Omega = \frac{1}{2} \epsilon_{A'B'}\sigma_A\sigma_B \dd x^{AA'}\wedge \dd x^{BB'}\wedge \langle \sigma \dd \sigma\rangle~.
 \end{align}
This coordinate system, and the dual sections, naturally define a basis for our homogeneous tensors, and we can expand our twistor space tensors in this basis $g=g_{\alpha_I} e^{\alpha_I}$. For example in the spin-three case
 \begin{align}
	g & = \left(g^{A_1A_2} \wedge (e_{A_1}\otimes e_{A_2})+g^{A_1}{}_{A'_2}\wedge (e_{A_1}\otimes e^{A'_2}) \right.\nn\\
		& \qquad \left. +\, g_{A'_1}{}^{A_2}\wedge (e^{A'_1}\otimes e_{A_2})+g_{A'_1A'_2}\wedge (e^{A'_1}\otimes e^{A'_2}) \right)\otimes \Omega ~,
 \end{align}
where the $g_{A_1A_2}$ etc. are $(0,1)$-forms of homogeneity $-6$
. By integrating over the fibres of twistor space we can now define space-time 2-forms $G^{A_1A_2}{}_{B_1 B_2}$, or more generally $G^{A_I}{}_{B_I}$, via
 \begin{align}
	G^{A_I}{}_{B_I}(x) =\int_{\Xi}\sigma_{B_I}g^{A_I}\wedge\Omega~.
 \end{align}
Motivated by the form of the anti-self-dual interactions in the linearized space-time action, 
 \begin{align}
	\int G^{A_I}{}_{B_I}\wedge G^{B_I}{}_{A_I} ~,
 \end{align}
as well as the interactions for conformal gravity and Yang-Mills, we consider the twistor space expression
 \begin{align}
	 \int_{\PTc\times_{\cal M} \PTc} \sum_{\abs{I} = 0}^\infty (\sigma_{1A_I} \sigma_{2B_I}) \, g^\Omega_1{}^{B_I}  \wedge  g^\Omega_2{}^{A_I} ~.
 \end{align}
Here the space $\PTc\times_{\cal M} \PTc$ is the space whose fibres over the manifold ${\cal M}$ are Cartesian products of the fibres of the individual twistor spaces ${\PTc}\rightarrow {\cal M}$, namely $\Xi_1\times\Xi_2 \simeq \mathbb{CP}^1\times \mathbb{CP}^1$, with homogeneous fibre coordinates $(\sigma_{1A}, \sigma_{2B})$. The fields $g_1$ and $g_2$ are $(0,1)$-forms depending on the respective fibre coordinates, while $\Omega_1$ and $ \Omega_2$ are the respective holomorphic volume forms. This action is constructed using the Penrose transform with respect to the background complex structure, and so at the linearised level it is obviously invariant under shift of $g$ by ${\bar \partial}\chi$ terms as they result in total derivative terms with respect to the fibre integration. However if we wish to include the effects of the deformation $f$, the equation satisfied by $g$, $\bar\partial_f g=0$, is modified. 

We account for this deformation by inserting the appropriate Green's function ${\bar\partial}^{-1}_F$ to propagate the fields in deformed twistor space along the fibres
 \begin{align}
 \label{eq:asd_act}
	S_{\rm a.s.d.} = \int_{\PTc\times_{\cal M} \PTc} \sum_{\abs{I} = 0}^\infty(\sigma_{1A_I} \sigma_{2B_I} )\, \left( \bar \partial\, {\bar\partial}_f^{-1}\smallborders_{\Xi_1}  \right) g^\Omega_1{}^{B_I}  \wedge  \left( \bar \partial\, {\bar\partial}_f^{-1}\smallborders_{\Xi_2}  \right)g^\Omega_2{}^{A_I}  ~.
 \end{align}

To define this action we must specify $\bar \partial\,{\bar\partial}_f^{-1}\smallborders_{\Xi}$ for which we use the assumption that the deformation is small and hence
 \begin{align}
	\bar \partial\,{\bar\partial}_f^{-1}\smallborders_{\Xi}\equiv\sum_{n=0}^\infty (\bar \partial^{-1}\! F\smallborders_{\Xi} )^n ~.
 \end{align}
We thus need to know how ${\bar\partial\smallborders}_\Xi^{-1}$ acts on holomorphic one-forms of homogeneity $n$, $\Omega(\mathbb{CP}^1, \mathcal{O}(n))$. This has been previously discussed in context of twistor actions, e.g. \cite{Mason:2008jy}, where it was shown that it can be expressed in terms of the Cauchy kernel. Only for the case of $n=-1$ this operation is uniquely defined. As $H^{0,1}(\mathbb{CP}^1, \mathcal{O}(n))$ is empty for $ n\geq -1$\,, every $k\in \Omega^{0,1}(\mathbb{CP}^1, \mathcal{O}(n))$ is exact, and so $k=\bar \partial \omega$ for some $\omega \in \Omega^0(\mathbb{CP}^1, \mathcal{O}(n))$. Additionally as $H^0(\mathbb{CP}^1, \mathcal{O}(n))$ is also empty for $n\leq-1$\,, there is no freedom in the definition of $\omega$. Concretely, suppose $\Xi \simeq \CP^1$ is parametrised by coordinates $\sigma_A$, we can define $\bar \partial^{-1}k$ by
 \begin{align}
{\bar\partial\smallborders}_\Xi^{-1} k(\sigma) & = \frac{1}{2\pi i}\int_\Xi \frac{\DD \sigma'}{\chirp{\sigma \sigma'}} \wedge k(\sigma')~.
\end{align}
When we consider forms with $n\geq 0$ we need to include additional factors to give the correct weight under coordinate rescalings, which can be done by using a reference spinor: $\left(\tfrac{ \chirp{\xi \sigma} }{ \chirp{\xi \sigma'}}\right)^{n+1}$. The arbitrariness in the choice of $\xi$ corresponding to the non-triviality of $H^0(\mathbb{CP}^1, \mathcal{O}(n))$ for $n > -1$ gives rise to a gauge freedom which should drop out of any physical observable. For $n < -1$ we find additional singularities which need to be specified. From the expression \eqref{eq:F_jet_connection} for the $(0,1)$-form $F$, we see that we have the following weights
 \begin{align}
	\left({\bar\partial}^{-1}\! F\smallborders_\Xi \right)(Z(\sigma)) 
		& = \frac{1}{2\pi i} \int_\Xi \frac{\DD \sigma'}{\chirp{\sigma \sigma'}} 
			\sum_{\abs{I},\abs{J} = 0}^\infty \left(\left(\tfrac{ \chirp{\xi \sigma} }{ \chirp{\xi \sigma'}}\right)^{\scriptscriptstyle \abs{I} + 1 + \abs{J}} 
						F^{\alpha_I;\beta_J}(Z(\sigma'))\T_{\alpha_I;\beta_J} \right.\nn\\
	&
	\kern+120pt \left. + \left(\tfrac{ \chirp{\xi \sigma} }{ \chirp{\xi \sigma'}}\right)^{\scriptscriptstyle \abs{I} + 1 - \abs{J}} 
						F^{\alpha_I}_{\beta_J}(Z(\sigma'))\T_{\alpha_I}^{\beta_J} \right) ~.
 \end{align}
The interaction of the anti-self-dual fields with the self-dual deformation is then encoded by the series
 \begin{align}
 \label{eq:ASD_exp}
	\bar \partial\,{\bar\partial}_F^{-1}\smallborders_{\Xi}\, g(Z) = g(Z) & +
		\inv{\bar\partial}\!\! \sum_{\abs{I},\abs{J} = 0}^\infty \left(
			F^{\alpha_I;\beta_J}(Z)\T_{\alpha_I;\beta_J} +
			F_{\beta_J}^{\alpha_I}(Z)\T^{\beta_J}_{\alpha_I} 
		\right)\smallborders_\Xi\, g(Z) + \dots 
 \end{align}
which when substituted into the action \eqref{eq:asd_act} generates interaction vertices involving all orders in the fields. These interactions
have an intricate structure which involves interactions between fields of different spins.

As mentioned, in principle one can find all tree-level amplitudes by starting with free plane-wave fields and iteratively solving the equations of motion. The exponentiated action evaluated on this classical solution is a generating functional for the amplitudes. To find the analogue of the gluon and graviton ${\rm MHV}$ amplitudes we can take the above interactions and expand the fields around their background values.

\section{Outlook}

One immediate generalisation of the current work is to include supersymmetry.	
Self-dual actions on super-twistor space have been previously considered for $\mathcal{N}=4$ SYM \cite{Witten:2003nn},
for $\mathcal{N}=4$ CSG \cite{Berkovits:2004jj} and for $\mathcal{N}=8$ Einstein gravity \cite{Mason:2007ct} while the twistor action for the full  $\mathcal{N}=4$ CSG was given in \cite{Adamo:2013tja}. From a geometrical perspective $\calN = 4$ supersymmetry is the most natural as it results in a Calabi-Yau super-twistor space, however it is by no means clear this theory is unique. At least in the spin two case, depending on the presence of certain additional global symmetries, the conformal supergravity theory has minimal and non-minimal versions and it is only the minimal version which contains Einstein supergravity as a truncation.

This raises the important question of what space-time theory our higher-spin twistor description actually corresponds to. At the level of the spectrum we have shown that the number of degrees of freedom matches with the number of on-shell states in the conformal higher spin theory described by Fradkin and Tseytlin \cite{Fradkin:1985am}.  Including the anti-self-dual interactions, a natural candidate is the four-dimensional case of Segal's conformal higher spin theory \cite{Segal:2002gd} which describes an infinite number of bosonic symmetric traceless tensor fields. In the unitary subsector we have seen that the spectrum of the linearised theory matches the spectrum of the Fronsdal theory \cite{Fronsdal:1978rb}. As Vasiliev's theory \cite{Vasiliev:1990en,Vasiliev:1992av,Vasiliev:2003ev} also reproduces the Fronsdal spectrum we may optimistically speculate that the full unitary subsector, given by $h_i$ and $\tilde h_i$, is related to the non-linear massless higher spin theory on anti-de Sitter space, however the absence of any scalar field in the twistor theory means that an exact matching would require some modifications. Nonetheless there are certain similarities, for example the higher-spin symmetry underlying the space-time theories can be understood as acting on jet spaces of fields and the unfolded formulation, see \cite{Shaynkman:2004vu} for the CHS case, involves a twistor space formulation with  some resemblance to the twistor spaces considered here. Of course to properly compare theories we must understand the structure of the interactions between the infinite tower of fields.

Our proposal \eqref{eq:asd_act} for the interaction terms of the anti-self-dual modes leads to an expansion about a given self-dual background
which will generate an infinite series of higher point vertices. Such expansions in ${\mathcal N}=4$ super-Yang-Mills and conformal gravity 
have led to efficient formalisms for computing observables, see \cite{Adamo:2011pv, Mason:2008jy, Mason:2009sa}. For example, fixing axial gauge for the the unitary truncation of the conformal gravity twistor action, Adamo and Mason \cite{Adamo:2013tja} were able to re-sum the resulting Feynman diagrams by using the matrix-tree theorem as in \cite{Feng:2012sy,Adamo:2012xe} to produce a formula for the de Sitter analogue of MHV amplitudes which reproduced Hodges' remarkable formula \cite{Hodges:2012ym} in limit of vanishing cosmological constant. 
The same calculation for the CHS theory, starting with the three-point all spin-$s$ ${\rm MHV}$ amplitude, may provide insight into the structure of the interactions of the theory. 

Finally, while we have focussed on the flat space-time background the twistor space actions are in principle valid for general self-dual space-times. Even at the quadratic level this of interest as  the space-time CHS kinetic operators are not known for general backgrounds, though there has been recent progress \cite{Metsaev:2014iwa, Nutma:2014pua, Beccaria:2014jxa}. In the context of one-loop checks of the correspondence between massless higher spin theories on anti-de Sitter space and vector model CFTs, e.g. \cite{Giombi:2013yva, Tseytlin:2013jya}, these are important objects as they are needed for computing  the canonical partition function on curved boundary manifolds
and twistor methods may provide an alternative method of calculation.

\section*{Acknowledgements}

We would like to thank Tim Adamo and Lionel Mason for many helpful discussions. This work was supported in part by Marie Curie Grant CIG-333851.

\appendix

\section{Twistor Space Geometry}
\label{sec:CurvedTwistorSpace}

Here we will review some background material regarding twistor theory that is necessary for our discussion.  We consider space-times corresponding to oriented four-dimensional Riemannian manifolds, which we denote by  ${\calM}$, with metric $g$ and possessing a spin structure. The unprimed and primed spinor bundles are denoted $\Spin^+$ and $\Spin^-$ with fiber coordinates $\alpha^A$ and $\mu^{A'}$, where $A, \ldots = 0,1$ and $A', \ldots =0,1$, and their dual bundles by  $\Spin^+{}^\ast$ and $\Spin^-{}^\ast$ with coordinates $\lambda_A$ and $\beta_{A'}$. In addition we have the skew-symmetric spinors $\epsilon_{AB}$, $\epsilon_{A'B'}$, $\epsilon^{AB}$, $\epsilon^{A'B'}$, which are defined such that the metric is given by 
 \begin{align}
	g_{ab} = \epsilon_{AB} \epsilon_{A'B'}~
 \end{align}
 where we denote four dimensional space-time indices by $a, b, \ldots =0,1,2,3$.
Furthermore $\epsilon^{AB}$ can be used to define the inner product 
 \begin{align}
	\langle \lambda \nu \rangle=\epsilon^{AB}\lambda_A\nu_B~, 
 \end{align}
and similarly $\chirm{\pi \mu} = \epsilon^{A'B'} \pi_{A'} \mu_{B'}$ as the inner product on the primed spinor bundle. 
In twistor theory it is standard to consider complex space-times where the spinor-bundles $\Spin^+$ and $\Spin^-$ are complex and unrelated to one another.
 This gives rise to new curvatures, in particular to a complex partner ${\tilde \Psi}_{A'B'C'D'}$ of the Weyl spinor ${\Psi}_{ABCD}$. Significantly, it is possible to have complex space-times for which ${\Psi}_{ABCD}=0$ but ${\tilde \Psi}_{A'B'C'D'}\neq 0$, that is geometries for which the anti-self-dual curvature vanishes but the self-dual one does not.

To construct the twistor space corresponding to a flat four-complex-dimensional space-time $\mathbb{C}M_4$ we consider the total space of the lower index un-primed spinor-bundle with points described by $(x^{AA'}, \sigma_A)$, where $x^{AA'}$ are the coordinates of $\mathbb{C}M_4$, using bi-spinor notation for coordinate indices. The projectivised spin-bundle $\Proj(\Spin^+{}^\ast)$ is a five-complex-dimensional manifold where $\sigma_A$ are interpreted as homogeneous coordinates on the $\CP^1$ fibres over $\mathbb{C}M_4$. We can define twistor space by projecting the $\Spin^+{}^\ast$ onto $\mathbb{T}$ with coordinates $Z^{\alpha}=(\lambda_A, \mu^{A'})$ by using the incidence relation
 \begin{align}
	\Spin^+{}^\ast  \ni (x^{AA'}, \sigma_A) \mapsto (\lambda_A, \mu^{A'})=(\sigma_A, x^{AA'}\sigma_A) \in {\bbT}~.
 \end{align}
Strictly speaking, this does not cover all of $\bbT$, and we should consider the conformal compactification of $\mathbb{C}M_4$, which implies a natural action of the conformal group.
The conformal group is isomorphic to $SU(2,2)$, which makes twistor space $\Tw \subset \Complex^4$ the representation space of the complex Weyl spinor representation of $\alg{su}(2,2)$. The same projection for the projectivised spin bundle defines projective twistor space $\PT \cong \CP^3$ which in  homogeneous coordinates is $Z^{\alpha}\sim t Z^\alpha$ for $t\in \mathbb{C}\backslash\{0\}$.

A curved twistor space $\Tc$ is a complex four-dimensional manifold with an Euler vector field $E$ and a non-vanishing holomorphic three-form $\Omega$ satisfying
 \begin{align}
	\Lie_E\Omega = 4 \Omega
	~~~{\rm and}~~~ 
	\iota(E)\Omega = 0 ~.
 \end{align} 
We can choose local homogeneous coordinates, $Z^\alpha$, $\alpha=0,1,2,3$, on $\Tc$ such that 
 \begin{align}
	E = Z^\alpha \frac{\partial}{\partial Z^\alpha} 
	~~~{\rm and}~~~
	\Omega = \frac{1}{6}\epsilon_{\alpha\beta\gamma\delta} Z^\alpha \dd Z^\beta \wedge \dd Z^\gamma \wedge \dd Z^\delta ~.
 \end{align}
 $\PTc$ corresponds to the space of orbits of $E$ in $\Tc$.
Curved projective twistor spaces $\PTc$ contain a four parameter family of compact holomorphic curves, $L_x$, with the topology of Riemann Spheres each of which has the same normal bundle as a $\CP^1$ in $\CP^3$. One identifies the points in the curved complex space, $x\in {\cal M}$, with these curves.

In order to consider real space-times, we must further define a reality structure. If we wish to choose ${\cal M}$ to be Lorentzian with signature $(1,3)$, we restrict to Hermitian $x^{AA'}$, and the primed and un-primed spinor bundles are related by conjugation. However, as this also relates the self-dual part of the Weyl curvature to the anti-self dual part the non-linear graviton construction can only be carried out in the conformally flat case. For real manifolds of definite signature --- for technical reasons the signature is in fact all negative --- there is an anti-linear conjugation on the spinors 
 \begin{align}
	\alpha^{A} \mapsto \hat \alpha^{A}~, 
	~~~\mu_{A'} \mapsto \hat \mu_{A'}~,
 \end{align}
such that $\hat{\hat{ \alpha}}^{A} =-{{\alpha}}^{A}$ and  $\hat{\hat{ \mu}}^{A'} =-{{\mu}}^{A'}$. For bi-spinors this conjugation is however involutative, and we define the real manifold to be the set of points satisfying $\hat{x}^{AA'}=x^{AA'}$. This induces a map on twistor space $Z^\alpha \mapsto {\hat Z}^\alpha$ that has no fixed points, however the lines in twistor space corresponding to the fixed space-time points are fixed lines. Consider the flat space-time case, ${\cal M}={\mathbb E}$; given two points on such a line $Z^\alpha$ and $\hat Z^\alpha$ we can define the projection ${\mathbb T}\rightarrow {\mathbb E}$ using the formula
 \begin{align}
	x^{AA'} = -i \frac{ \hat{\lambda}^A \mu^{A'} - \lambda^A \hat{\mu}^{A'} }{\langle \lambda \hat{\lambda}\rangle}~.
 \end{align}
In this case the unprimed spinor bundle, which is now an eight-dimensional real manifold, and twistor space $\mathbb{T}$ can be identified, and similarly for the projective spinor bundle and projective twistor space. 
 
This is true in the general case where the curved twistor space ${\Tc}({\cal M})$ is identified with the unprimed spinor bundle over ${\cal M}$ with non-holomorphic coordinates $(x^{AA'}, \sigma_A)$ where $ \sigma_A$ is a spinor at $x^{AA'}\in {\cal M}$. There is an almost complex structure on $\Tc$ such that the space of $(0,1)$-tangent vectors at $(x^{AA'}, \sigma_A)$ is spanned by 
 \begin{align}
	\hat{V}_{A'} = -\frac{i}{\langle \sigma \hat{\sigma}\rangle} \sigma^A\nabla_{AA'}
	~~~{\rm and}~~~ 
	\hat{V}^A=\hat{\partial}^A=\frac{\partial}{\partial \hat{\sigma}_A}~.
 \end{align}
Projective twistor space $\PTc({\cal M})$ corresponds to the projective unprimed spinor bundle over ${\cal M}$ where $\sigma_A$ are now the  homogeneous coordinates of the ${\CP}^1$ fibre. The almost complex structure on ${\Tc}({\cal M})$ reduces to $\PTc({\cal M})$ as the tangent space of projective twistor space can be found by factoring out the fields $E=\sigma_A\partial^A$ and ${\hat E}=\hat{\sigma}_A\hat{\partial}^A$\,. As shown by Atiyah \textit{et al}, \cite{atiyah1978self}, the almost complex structure is integrable if and only if ${\cal M}$ has vanishing anti-self-dual Weyl curvature $\Psi_{ABCD}=0$\,. This almost complex structure reduces to a complex structure on $\PTc({\cal M})$ as the vector $E$ is holomorphic.

For a generic self-dual manifold we cannot generally define global twistors. However, we can define at each point $x\in {\cal M}$ a local twistor $Z^{\underline \alpha}$ which for a given metric $g$ is represented by a pair of spinors $(\lambda_A, \pi^{A'})$ which transform as 
 \begin{align}
 	\begin{aligned}
	\tilde \lambda_A & = \lambda_A - i \Upsilon_{AA'} \pi^{A'}\\
	{\rm and} ~~~
	\tilde \pi^{A'} & = \pi^{A'}
	\end{aligned}
 \end{align}
under the Weyl transformation $g\mapsto \tilde g=\Omega^2 g$ with $\Upsilon_{AA'}=\nabla_{AA'}\log \Omega$\,. We denote the corresponding rank-four local twistor bundle over ${\cal M}$ by $\mathbb{LT}$. Pulling back a section of $\mathbb{LT}$ given by the spinor fields $(\lambda_A, \pi^{A'})$, we can define a (1,0)-vector field on ${\mathbb T}({\cal M})$ by 
 \begin{align}
\label{eq:holvec}
	T=\lambda_A(x) V^A+ \pi^{A'}(x) V_{A'}~,
 \end{align}
where $V^A=\partial^A$ and $V_{A'}=\tfrac{-i}{\langle \sigma \hat{\sigma}\rangle} \hat{\sigma}^A\nabla_{A'A}$\,. One can show, see e.g. \cite{woodhouse1985real}, that such vectors are holomorphic if and only if they are parallel under local twistor transport. That is to say when they satisfy the conditions
 \begin{align}
 	\begin{aligned}
	& \nabla_{BB'}\pi^{A'}+i \lambda_A \epsilon_{B'}{}^{A'}=0 \\
	{\rm and}~~~ 
	& \nabla_{BB'} \lambda_{A}-i (\Phi_{ABA'B'} -\Lambda \epsilon_{AB}\epsilon_{A'B'})\pi^{A'}=0~,
	\end{aligned}
 \end{align}
where $\Phi_{ABA'B'}$ is proportional to the trace-free Ricci tensor and $\Lambda$ to the Ricci scalar. In the conformally flat space case these equations have a four-complex-parameter family of solutions given by 
 \begin{align}
	\pi^{A'} = \mu^{A'}_0 + i \lambda_A x^{A A'}
 \end{align}
for constant $\mu^{A'}_0$ and $\lambda_A$. We can thus identify the solution space of the twistor equation with flat twistor space $\mathbb{T}\simeq \mathbb{C}^4$. Given a holomorphic field, $T$, of the form \eqref{eq:holvec} on $\mathbb{T}$ we can choose linear coordinates $Z^{\alpha}$ such that 
 \begin{align}
	T = T^\alpha\frac{\partial}{\partial Z^\alpha}
 \end{align}
with $T^\alpha$ being constant.

\renewcommand{\bibname}{References}
 \bibliographystyle{utcaps} 
\bibliography{ampsCG}

\end{document}